\documentclass[fleqn,10pt]{wlscirep}

\usepackage[utf8]{inputenc}
\usepackage[T1]{fontenc}

\graphicspath{{./figures/}}
\DeclareGraphicsExtensions{.PDF,.eps,.jpg,.png,.pdf}
\usepackage{amsmath}
\usepackage{xcolor, colortbl}
\usepackage[caption=false]{subfig}
\usepackage{soul}
\usepackage{bm}
\usepackage{amsfonts}

\usepackage{dsfont}
\usepackage{lipsum}
\usepackage{booktabs}
\usepackage{nicematrix}
\DeclareMathOperator*{\argmax}{arg\,max}

\hypersetup{colorlinks=true,linkcolor=black,anchorcolor=black,citecolor=black,filecolor=black,menucolor=black,runcolor=black,urlcolor=black}

\makeatletter
\newcommand\remembertext[2]{% #1 is a key, #2 is the text
  \immediate\write\@auxout{\unexpanded{\global\long\@namedef{mytext@#1}{{\color{blue} #2}}}}%
  {\color{blue} #2}%
}
\newcommand\remembertextnocolor[2]{% #1 is a key, #2 is the text
  \immediate\write\@auxout{\unexpanded{\global\long\@namedef{mytext@#1}{#2}}}%
  #2%
}
\newcommand\recalltext[1]{%
  \ifcsname mytext@#1\endcsname
    \@nameuse{mytext@#1}%
  \else
    ``??''
  \fi
}
\makeatother
\title{Automatic Detection and Annotation of Sperm Whale Codas}

\author[1,+]{Guy Gubnitsky}
\author[2,+]{Yaly Mevorach}
\author[3,4,+]{Shane Gero}
\author[5,+]{David F. Gruber}
\author[1,6,*,+]{Roee Diamant}
\affil[1]{University of Haifa, Department of Marine Technology, Haifa, 3498838, Israel}
\affil[2]{University of Haifa, Morris Kahn Marine Research Station, Department of Marine Biology, 3498838, Israel}
\affil[3]{Carleton University, Department
of Biology, Ottawa, Ontario, K1S 5B6, Canada}
\affil[4]{The Dominica Sperm Whale Project, Roseau, Dominica}
\affil[5]{City University of New York, Baruch College, New York, NY, 10010, USA}
\affil[6]{University of Zagreb, Faculty of Electrical Engineering and Computing, Zagreb, Unska 3, Croatia}
\affil[+]{Project CETI, New York, NY 10003, USA}
\affil[*]{roee.d@univ.haifa.ac.il}


\begin{abstract}
    A key technology in sperm whale (\textit{Physeter macrocephalus}) monitoring is the identification of sperm whale communication signals, known as \textit{codas}. In this paper we present the first automatic coda detector and annotator. The main innovation in our detector is graph-based clustering, which utilizes the expected similarity between the clicks that make up the coda. Results show detection and accurate annotation at low signal-to-noise ratios, separation between codas and echolocation clicks, and discrimination between codas from simultaneously emitting whales. Using this automatic annotator, insights into the characterization of sperm whale communication are presented. The results include new types of coda signals, analyzes of the distribution of coda types among different whales and for different years, and evidence for synchronization between communicating whales in terms of coda type and coda transmission time. These results indicate a high degree of complexity in the communication system of this cetacean species. To ensure traceability, we share the implementation code of our coda detector.
\end{abstract}

\begin{document}

\flushbottom
\maketitle
\thispagestyle{empty}

\section*{Introduction}
\label{sec1}
In recent years, considerable efforts has been made to record and capture the bioacoustics of sperm whale (\textit{Physeter macrocephalus}). For example, Project CETI (Cetacean Translation Initiative), founded in 2020, is an interdisciplinary research initiative that uses advanced machine learning and cutting-edge robotics to better understand sperm whale communication~\cite{andreas2021cetacean}. The backbone of this effort is custom-built passive bioacoustic arrays covering a 20×20 kilometer area where these whale families reside (collecting over 30 TB/month), in conjunction with robotic acoustic and video tags on the whales, underwater gliders and aerial drones to augment the rich contextual communication data. This makes it possible to train a complete generative language model for cetacean communication ~\cite{goldwasser2023}.

More broadly, in the face of biodiversity loss and the climate crisis, armed with rapidly miniaturizing hardware to record wildlife and wild spaces, scientists are challenged not by collecting increasing large, and continuous amounts of data; but by devoting the time and development of the software tools to curate and interpret it (see \cite{muller2023soundscapes}). Automated species detection and annotation of vocalizations is a critical step to characterizing and monitoring our changing ecosystem, and this is even more true in the worlds oceans. Further, it allows for rapid assessment of previously data depauperate areas. In this case, given the importance of culture in conservation \cite{brakes2019animal, brakes2021deepening}, and that the cultural population structure of sperm whale clans is defined by acoustic repertoire \cite{rendell2003vocal, hersh2021method}, an automatic detection and annotation of sperm whales codas will allow for the definition of clans present in previously unrecorded waters using passive acoustics, and thus provide not only a tool for understanding this species' communication system, but also for their applied conservation and management.

\subsection{Current Approaches}

Sperm whales are among the most acoustically active toothed whales, making them attractive species for passive acoustic monitoring (PAM). Many methods have been proposed to detect and classify sperm whale clicks~\cite{sanchez2010efficient,roch2008comparison,lohrasbipeydeh2014adaptive,kandia2006detection,kandia2008phase,beslin2018automatic}. However, existing techniques focus on the detection and classification of echolocation clicks, which whales use to navigate the darkness of the deep sea and hunt squid prey. However, when individuals communicate with each other, sperm whales use an acoustically distinct type of click \cite{madsen2002sperm,Mohl2003}, which are produced in short sequences with stereotyped rhythm and tempo called 'codas' \cite{watkins1977sperm, schulz2008overlapping, weilgart1993coda}. Codas last less than two seconds and have been characterized by the varying number of constituent clicks and the intervals between them. Codas are typically generated in multiparty conversations that can last from 10 seconds to over half an hour. %Each click within the coda presents as a sequence of evenly spaced pulses which is the result of sound reflection within the spermaceti organ.

Coda recognition is usually done by automatic click detection and subsequent manual annotation to distinguish coda clicks from echolocation clicks, to separate vocalizers and to 'link' clicks to the same coda. Existing manual annotation systems include custom Matlab scripts \cite{rendell2004shared,rendell2003comparing}, the Rainbow Click software developed by the International Fund for Animal Welfare \cite{schulz2008overlapping, gero2016individual} and more recently CodaSorter \cite{Vachon2022oceanNomads,hersh2021method}.%, a custom software developed by the Marine Bioacoustics Lab at Aarhus University.
However, while all of these programs provided a user interface for manual annotation of codas by users based on feature extractions from the clicks (e.g. interpulse interval, angle of arrival, and spectra content of the clicks), none of these programs provided automatic detection of codas, classification of echolocation and coda clicks, or the ability to automatically separate sources and automatically annotate individual clicks as belonging to the same coda. While Beslin et al. \cite{beslin2018automatic} performed classification between echolocation and coda clicks, this solution omits codas as "bad" echolocation. As a result, there are no automated approaches for detecting coda clicks yet. This may be due in part to the difficulty of separating sources between vocalizers, as codas from multiple whales often overlap during exchanges \cite{schulz2008overlapping, sharma2023contextual}.

In this work an automated detector for codas is offered. The detector is intended for both presence detection and characterization of codas, while separating overlapping sources. The detector is designed for a wide dynamic range to enable recordings from both an acoustic tag attached to a sperm whale (near field) and an acoustic recorder deployed from a boat or mooring (far field). The complexity of the detector is low, so that real-time operation is possible. Offline analysis of data from acoustic tags that were not previously processed revealed new insights into the synchronization of coda characterization between pairs of communicating whales as well as new types of coda signals.

\subsection{Preliminaries: characteristics of the Coda signal}

The codas of sperm whales consist of patterns of broadband, impulsive signals called \textit{clicks}, which are characterized by a multi-pulsed structure \cite{Mohl2003}. The first impulse is generated by forcing air through the whale's phonic lips. The generated sound is reflected by the air sacs at the anterior and posterior ends of the spermaceti organ. The result is a series of pulses of decreasing amplitude that follow each other at equal time intervals; a product of the transit time between the nasal air sacs in both directions. The two most important features for distinguishing between coda and echolocation clicks are the decay rate between pulses within the two click types, which is probably the result of different signaling pathways in the whale's nose \cite{huggenberger16}, and their rhythmic structure. The echolocation clicks are produced in long series with a periodic pattern with characteristic inter-click intervals (ICIs) which are a function of the whale's search range~\cite{madsen2002sperm}. In contrast, codas are sequences of clicks with stereotyped rhythm and tempo \cite{watkins1977sperm}, which are thought to serve communication during exchanges between multiple individuals \cite{schulz2008overlapping}, and may serve social recognition and identity~\cite{gero2016individual, hersh2022evidence, rendell2003vocal}.

\begin{figure}[]
	\centerline{		\includegraphics[width=130mm,angle=0]{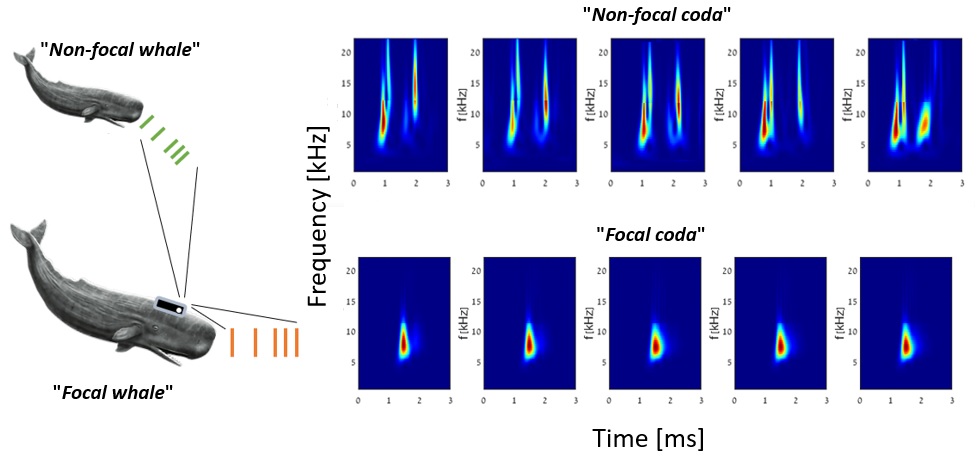}}
	\caption{Illustration of a scenario with a focal and a non-focal whale emitting 5-click codas. An example of a super-resolution spectrogram of individual clicks from a real tag recording of the codas of the focal and non-focal whale are shown in the top right and bottom right panels, respectively. A clear distortion of the structure of the clicks can be seen in the coda received from a distance. However, the similarity between the clicks is retained.}
	\label{Challenges}
\end{figure}

One challenge in recognizing codas is managing distortions caused by multipath interference. These lead to high variability in coda features, making it difficult to distinguish between a coda, a sequence of echolocation clicks or other interfering transients such as snapping shrimps or shipping cavitation noises. These challenges are highlighted by the example in Fig.~\ref{Challenges}. The figure shows recordings of two codas from a Dtag\footnote{A Dtag is an acoustic tag that is attached to the skin of the whale with a suction cup to measure acoustic recordings as well as depth and temperature information \cite{DtagPaper}} attached to a sperm whale. The bottom panel shows a coda originating from the tagged whale (the focal whale) and the top panel shows a coda received from a distant whale. The codas are represented by super-resolution spectrograms~\cite{moca2021time}. While the coda of the focal whale shows a stable tempo-spectral structure, we observe that the coda received from the distant whale is distorted.

While the coda's clicks may be distorted~\cite{schulz2009off, jacobs2024active}, it is clear from Fig.~\ref{Challenges} that this distortion is typical for all clicks. This is due to the fact that the same acoustic channel is used for all coda clicks. Such similarity is also evident in the clicks' inter-pulse intervals (IPI), which are strongly correlated with the size of the whale's spermaceti organ~\cite{madsen2002sperm,mohl1981}. This observation motivates us to build our coda detector on a similarity metric to identify sequences of statistically related clicks. The result is a clustering scheme that, given a set of detected clicks (e.g., the method in \cite{gubnitsky2023detecting}), separates groups of clicks with similar structure based on a likelihood metric. To distinguish between echolocation clicks and codas, we constrain the clustering solution to signals with a large multipulse structure and a resonant frequency below a threshold.

\section{Results}\label{sec2}

\subsection{Description of the Testbed}

Fewer than 600 individuals have been identified in the sperm whale community in the Eastern Caribbean \cite{VachonAbundance, Gero2007}. These are divided into at least two, but perhaps three cultural clans \cite{Vachon2022oceanNomads}. Matrilineal social units of sperm whales associate with other units of whales which share a similar dialect of codas, creating a culture-based population structure known as \textit{clans}. In the Eastern Caribbean, there are more members of the EC1 clan ($>$ 200 whales) than members of the EC2 clan ($<$ 200 whales) \cite{VachonAbundance}. Known social units of female and immature sperm whales \cite{Gero2014} were located and tracked in an area covering approximately 2000 $\mathrm{km}^2$ along the entire west coast of the island of Dominica (N15.30 W61.40).

Our dataset includes both near-field data from 42 tags deployed on 25 different individuals in 11 different social units, with photo-identification to match the tagged whale, and far-field data by hydrophone deployed from a boat. Preliminary manual annotations for near-far data between 2014-2016 yielded a set of 3948 codas, while a set of 4930 codas was obtained for far-field data between 2005-2012. Additional annotations were supported by our automatic detector to analyze 14.9~GB of near-field data collected in 2018, yielding a further set of 843 codas. Detector-assisted annotations also enabled the processing of 25~GB of far-field data, yielding 727 codas. For the assessment of false alarms, noise segments were used, which were divided into two categories: 1) four hours of raw acoustic signals that were manually verified to include ambient noise, ship noise, whistles and clicks from other marine mammals, and 2) three hours of echolocation clicks from several sperm whales, which were used to investigate the detector's ability to distinguish between codas and echolocation clicks. %In addition, we used **** hours of manually verified recordings of ambient noise collected at the THEMO observatory~\cite{diamant2018themo}, approximately 11~km west of the northern coast of Israel, which were used for an extensive statistical evaluation of the false alarm.

\subsection{Detection and Annotation Results}

\begin{figure}[t]
	\centering
	\subfloat[Noise including echolocation clicks.\label{fig:ROC_a}]{\includegraphics[width=0.48\columnwidth]{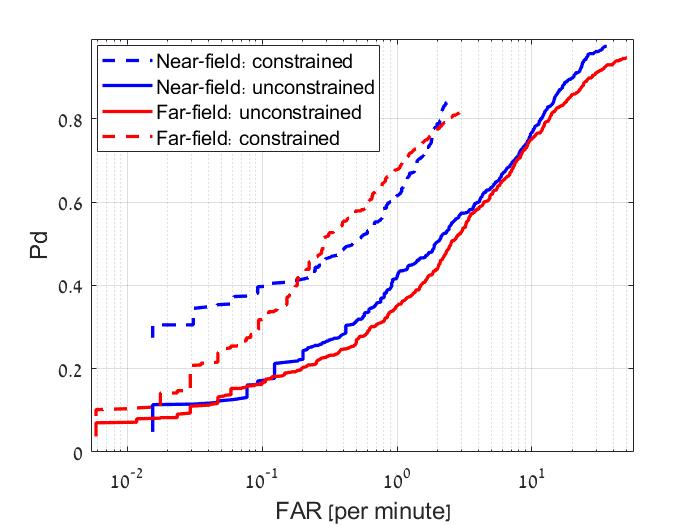}}
	\hspace{2mm}
 	\subfloat[Noise do not include echolocation clicks.\label{fig:ROC_b}]{\includegraphics[width=0.48\columnwidth]{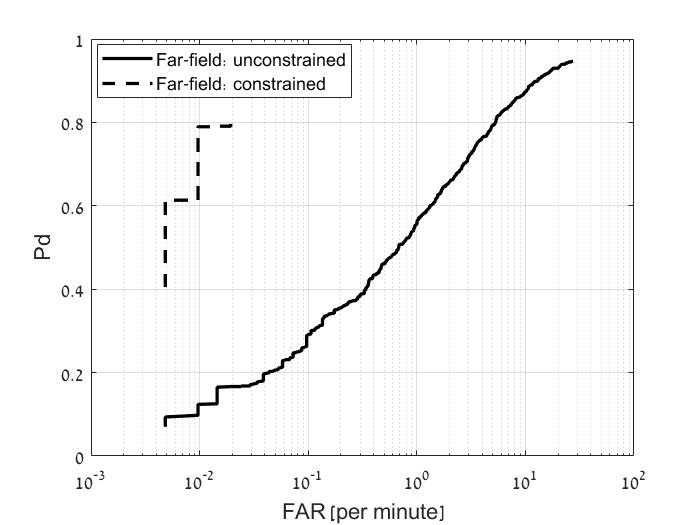}}
	\caption{ROC curve, evaluated for near- and far-field data. The solid lines refer to the unconstrained detection mode, while the lines for the constrained mode are dashed. False alarm rate evaluated for noise data including a) sperm whale echolocation clicks, b) ambient noise without echolocation clicks (only far-field data).}
    \label{fig:ROC}
\end{figure} 

\begin{figure}[]
	\centerline{		\includegraphics[width=100mm,angle=0]{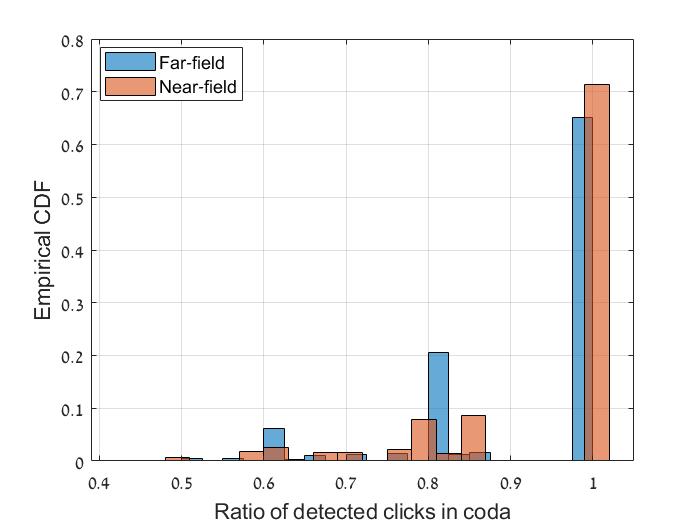}}
	\caption{Cumulative distribution function (CDF) of the ratio of the number of clicks detected in the coda in the near- and far-field datasets using constrained detection mode.}
	\label{fig: rank ration}
\end{figure}

The detection performance is analyzed in Fig.~\ref{fig:ROC} in terms of the Receiver Operating Characteristic (ROC) to investigate the trade-off between the detection rate (Pd) and the false alarm rate (FAR). The former is defined by the rate of true-positive cases verified by the manual annotations, and the latter by the rate of false-positive cases per minute. Pd and FAR were calculated based on manual annotations. As explained in Section~\ref{sec:method}, the detector applies a maximum likelihood approach, which is based on a distribution analysis of legacy database. To avoid overfitting, these distributions were evaluated from our near-field database from 2014-2016 and our far-field database from 2005-2012, while the performance is shown for the most recent data collection, i.e. for the near-field from 2018 and the far-field from 2022-2023.

The performance is compared for noise with echolocation clicks (Fig.~\ref{fig:ROC_a}) and for noise without echolocation clicks (Fig.~\ref{fig:ROC_b}). The detection results are shown for both far-field and near-field data\footnote{The results for noise without echolocation clicks are for far-field only, as the data from tags almost always contained echolocation clicks.}, and for constrained and unconstrained detection (see details in Sections~\ref{sec: Clustering} and ~\ref{Interpulse analysis}). The results show that the FAR reduces by an order of magnitude when using constraints, at the cost of a slight decrease in the detection rate. For the echolocation noise (Fig.~\ref{fig:ROC_a}), no significant difference in performance is observed between the near-field and far-field datasets. From this, we conclude that our detector can handle a low signal-to-noise ratio (SNR) well. Comparing the results for the two noise categories, we find that the FAR is twice as high for detection without constraints when the noise contains echolocation clicks, which is the harder case. Since the multipulse and resonant frequency constraints also allow the separation of noise transients, this difference is magnified when considering the performance differences between the constraint detectors.  

%of a constrained and an unconstrained solution are compared. The constraints include the limitation of the detector by the number of identified interpulses of the clicks denoted by $\mathcal{P}$, and by the limitation of the maximum allowed resonance frequency denoted by $f_r$, which are explained in more detail in Sec.~\ref{sec: Clustering} and ~\ref{Interpulse analysis}. Fig.~\ref{fig: ROC} compares detector performance over tag and remote recordings, while false detection is evaluated over an environment filled with echolocation clicks. The results show a reduction in FAR of about an order of magnitude for both tag and remote data at the cost of a 0.1 reduction in detection rate when the proposed constraints are used. There are no significant performance differences between remote and tag measurements, indicating that our detector is robust even at low SNR. A slight inferiority of the constrained detector can be observed in the tag data. The reason for this lies in the larger number of cases where the clicks overlap. Since the duration of the focal whale's clicks is much longer in the tag recordings (e.g., the example in Fig.~\ref{Definitions}), the probability of overlap increases considerably and impact the detection performance. 

In Table.~\ref{Results- FAR}, we examine the effects of each constraint for both cases of noise categories for the far-field dataset. Two detection thresholds are selected by a trade-off of 1) Pd=0.18 and FAR=0.12 per min and 2) Pd=0.4 and FAR=0.33 per min. To investigate the effects of the clustering constraints, we show the results without constraints (None), with the multipulse constraint only ($\mathcal{P}$), with the resonant frequency constraint only ($f_r$) and with both constraints ($\mathcal{P}$ \& $f_r$). In both cases of noise categories, we find that the multipulse constraint is most dominant in distinguishing coda clicks from echolocation clicks and noise transients. The resonant frequency constraint also contributes, but less. We explain this result by the effect of the channel impulse response, which distorts the spectrum of the far-field signal and allows less separation by spectra analysis. For the more common case of ''general noise'' in far-field data, we observe a very low FAR for constraint detection.

\begin{table}[t]
\caption{False alarm rate (FAR) per minute. Results are given for far-field data with detection thresholds set according to detection rate of \{0.18,0.4\} and a false alarm rate of \{0.12,0.33\}. The results are analyzed for two noise categories: without echolocation clicks (general noise) and with echolocation clicks (echolocation clicks).}
\begin{tabular}{lcccccccl}\toprule
& \multicolumn{4}{c}{General Noise} & \multicolumn{4}{c}{Echolocation Clicks}
 \\\cmidrule(lr){2-5}\cmidrule(lr){6-9}
      Constraint applied      & $\mathcal{P}$  & $f_r$ & $\mathcal{P}$ \& $f_r$ & None  & $\mathcal{P}$   &  $f_r$ & $\mathcal{P}$ \& $f_r$ & None\\\midrule
Target: Pd=0.18 \& FAR=0.12   & 0.0097  & 0.024 &  ~~0  & 0.034 & 0.035 & 0.082 & ~0.023  & 0.12 \\
Target: Pd=0.4 \& FAR=0.33   & 0.0097  & 0.275 &  ~~0  & 0.33 & 0.199 & 0.987 & ~0.152  & 1.5 \\
 \\\bottomrule
\end{tabular}
\label{Results- FAR}
\end{table}

%of the detector for two noise conditions: general noise and interference of sperm whale echolocation clicks, and we compare the contribution of the proposed constraints. The results at thresholds set according to the remote dataset for a detection rate of 0.8 show a high resilience to general noise when the two constraints are combined.  For both noise environments, the constraints reduce the false detection rate, with the P constraint taking precedence over the spectral constraint fr and the combination of both over using each constraint separately.

Next, we examine the accuracy in the automatic annotation of all clicks within the coda. This attribute is important to define the type of coda, which in turn is determined by the number of clicks and the rhythm and tempo of the ICI pattern~\cite{sharma2023contextual, weilgart1993coda}. Especially in the presence of echolocation clicks or in the frequent case of overlapping codas, a true coda click can easily be disqualified as an echolocation. Fig.~\ref{fig: rank ration} shows the cumulative distribution function (CDF) of the ratio between the number of identified clicks and the actual clicks in the coda. We compare the performance for the far-field and near-field datasets. Identification of all clicks in the coda is achieved in about 70\% of cases, while in another 10-20\% of cases 80\% of clicks are identified. The results show no significant difference between the near-field and far-field datasets. Although the accuracy level in detecting the coda clicks is promising, we note that misidentification of a single click may lead to misclassification of the coda type. We leave the improvement of this result to future work.

\subsection{Communication Characterization: findings from the coda annotator}

\begin{figure}[]
	\centerline{		\includegraphics[width=130mm,angle=0]{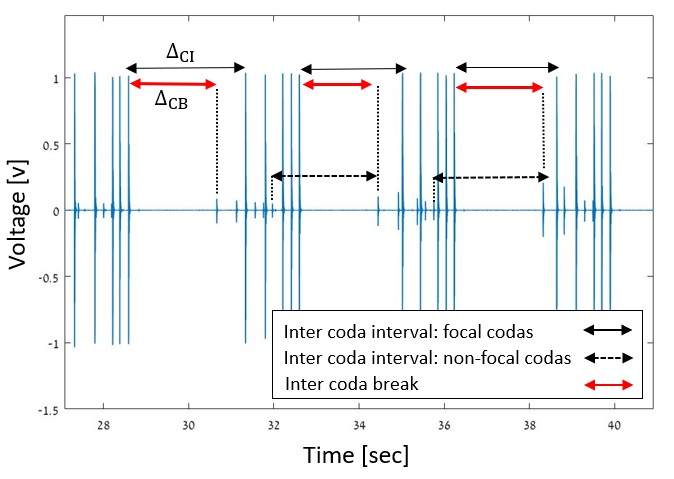}}
	\caption{An example of 13 seconds from a near-field recording showing a dyadic coda exchange between a focal and a non-focal whale. The Inter Coda Interval of the focal and non-focal whale are indicated by black solid and dashed arrows, respectively. The Inter Coda Break is indicated by red solid arrows.}
	\label{Definitions}
\end{figure}

The ability of the developed automatic detector and annotator to provide large datasets of coda was used to investigate the characteristics of coda signals towards the understanding of sperm whale communication. To eliminate bias in our analysis, we filter our coda database to use only codas from the near field, which allows determination of the identity of the focal whale. Further, since no classifier is currently proposed to distinguish between codas of individual whales, the following analysis considers dyadic social interactions by codas involving only two whales. To that end, we only consider coda pairs (see definition bellow) between a focal and a non-focal whale. These cases are identified by considering pairs of codas, detected within a buffer of 7~sec and where there is a significant gap in both codas' amplitude. Other codas that contained a single coda with no detected response or codas of more than two whales were discarded. An example of a dyadic interaction of 3 coda pairs is shown in Fig.~\ref{Definitions} by the red arrow marks. This process resulted in 635 coda pairs used for the below analysis.

Our main findings from the operation of the automatic coda annotator are evidence for synchronization between the focal and non-focal whales. This was established by examining the time delay between successive codas and between the clicks of the two codas. This type of correspondence arguably corresponds to the synchronization required for a stable telecommunication session. In particular, we note that the structure of the coda can be considered as a communication modulation signal. The time interval between codas can then be referred to as the communication baud rate, while variations between the click delays (ICI) of the codas can reflect the information encoding to obtain the channel's capacity outage. Another finding is the discovery of two new coda structures. This was revealed by comparing the extracted coda features with those of codas available in the literature. To describe these two findings, the following definitions are used, and examples of each of these structures are given in Fig.~\ref{Definitions}.

\begin{itemize}
\label{sec: definitions}
    \item \textit{Coda Type}: Categorical coda representation obtained by clustering codas based on their ICI vectors~\cite{gero2016individual}. 
    \item \textit{Dyadic coda exchange}: Period in which codas are made by a pair of whales. 
    %\item \textit{Meter}: A given whale produces codas with a fairly isochronous periodicity in which speakers make codas with an regular interval \cite{schulz2008overlapping, sharma2023contextual}. The time interval, $\Delta_{M}$, between each successive pair of calls of a particular whale. 
    \item \textit{Inter-coda Interval}: A time interval, $\Delta_{CI}$, between a consecutive pair of calls of a particular whale. As marked in Fig.~\ref{Definitions}, the inter-coda interval for a given whale is measured by the time delay between the last click of a coda and the first click of a subsequent coda of the same whale.
 \item \textit{Coda pair}: Two consecutive codas originating from different whales in a dyadic exchange within a buffer of 7~sec, such that the first click of one coda is received before the last click of the second coda.
 \item \textit{Inter-coda Break}: The time interval, $\Delta_{CB}$, between a coda pair. As marked in Fig.~\ref{Definitions}, the Inter-coda Break is defined by the time delay between the last click of one coda and the start of the first click of the interlocutor's next coda.
\end{itemize}

\subsubsection{Modulation Signal: Analysis of Coda Type Distribution}

A coda type is characterized by the number of clicks it consists of (between 3 and 10) and by the sequence of time delays between the individual clicks. Examples of this are a sequence of 5 clicks in which the first three clicks are evenly spaced and at a different distance from each other than the last two clicks (type 1+1+3), or a sequence of 8 clicks with a gradually increasing or decreasing distance (type 8i or 8D, respectively). To date, 25 coda types have been identified \cite{gero2016individual}. These were determined by a clustering result applied to annotated ICI vectors of codas \cite{gero2016individual,Gero2016-social}.

\begin{figure}[]
	\centerline{\includegraphics[width=\textwidth]{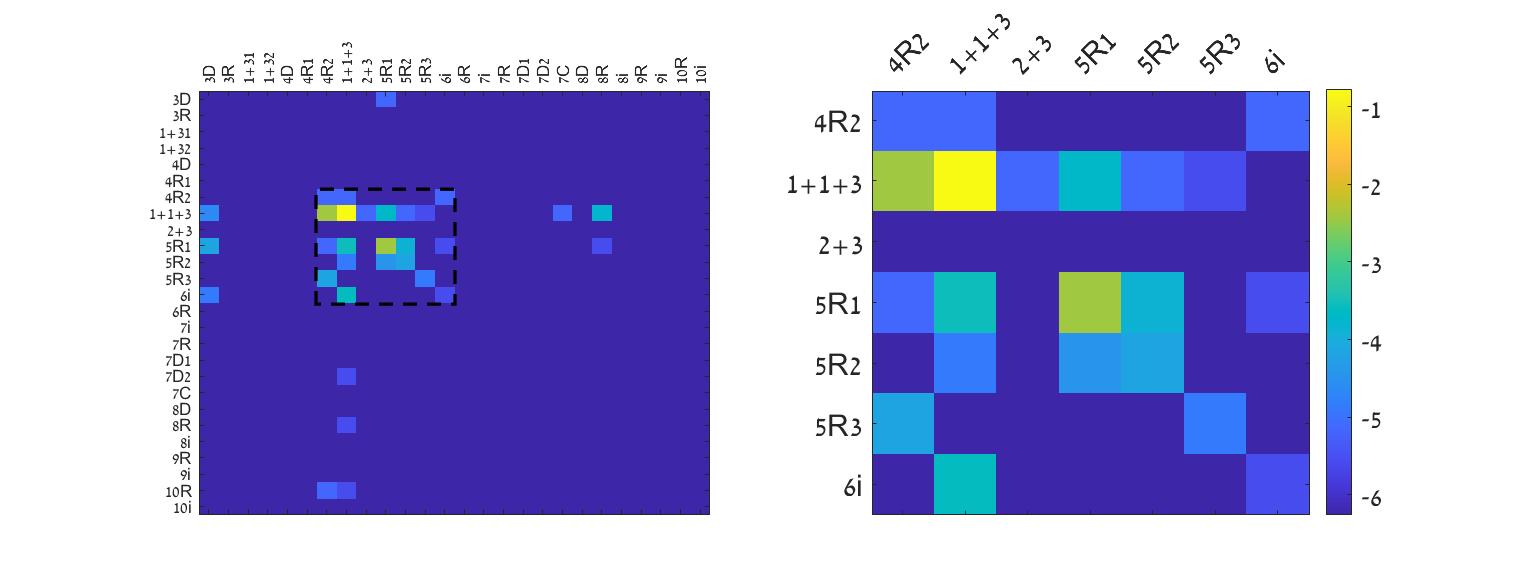}}
	\caption{Analysis of the distribution of coda types for all dyadic coda exchanges found. The left panel shows the distribution matrix of all pairs of coda types, with each bin representing a particular coda type from the focal whale (signal) and the subsequent coda type of the non-focal whale (response). The right panel is zoomed in on the most active coda types. The bins' color represents the probability of occurrence on a logarithmic scale.}
	\label{Time lag dependencies}
\end{figure}
The analysis in Fig.~\ref{Time lag dependencies} shows the distribution of codas for all considered coda pairs found during the periods of dyadic coda exchange. The results are presented as a matrix for the distribution of a 'signal and response' (S\&R) session between the focal and non-focal whale. Without loss of generality, the coda type of the focal whale in the matrix's columns is considered here as 'signal' (S) and the coda type of the non-focal whale (hereafter referred to as the \textit{Interlocutor}) in the matrix's columns is considered the 'response' (R). The bin's color corresponds to the probability (on a logarithmic scale) that a particular focal's coda type is observed before a particular interlocutor's coda type. The results show that the '1+1+3' type is the most typical coda type for both the signaler and the responder. Furthermore, the variation in the coda types is not large and the highest concentration is between types '1+1+3' and itself, between types '1+1+3' and '4R2' and between type '5R1' and itself.

\iffalse 
\begin{figure*}[t!]
    \centering
    \begin{subfigure}[t]{0.95\textwidth}
        \centering
        \includegraphics[height=1.5in]{Figures/QA_table.png}
        \caption{Q\&A tables. On the left is the table structure; the table's columns contain the coda types recorded from the focal whale and the rows contain those recorded from his interlocutor: the non-focal whale. On the right is the table layout. Each cell color represents the frequency of Q\&A occurrences of certain coda types.}
        \label{communication sessions- all whales}
    \end{subfigure}

\hfill
\begin{subfigure}{0.48\textwidth}
    \includegraphics[width=\textwidth]{Figures/Type_to_type_per_year_per_speaker.png}
    \caption{ Q\&A per speaker per year}
    \label{communication sessions- per whale}
\end{subfigure}
\hfill
\begin{subfigure}{0.48\textwidth}
    \includegraphics[width=\textwidth]{Figures/Type_to_type_per_year.png}
    \caption{Q\&A of several whales per year}
    \label{communication sessions- per year}
\end{subfigure}

    \caption{}
    \label{communication sessions}
\end{figure*}
\fi

\subsubsection{The Baud-rate: Time Delay Dependencies in Whale Codas}

\begin{figure}[t]
	\centering
	\subfloat[Probability density of $\Delta_{\mathrm{CB}}-\Delta_{\mathrm{CI}}$ for three focal whales, 'Atwood','Fork' and 'Pinchy'. \label{QA per whale}]{\includegraphics[width=0.48\columnwidth]{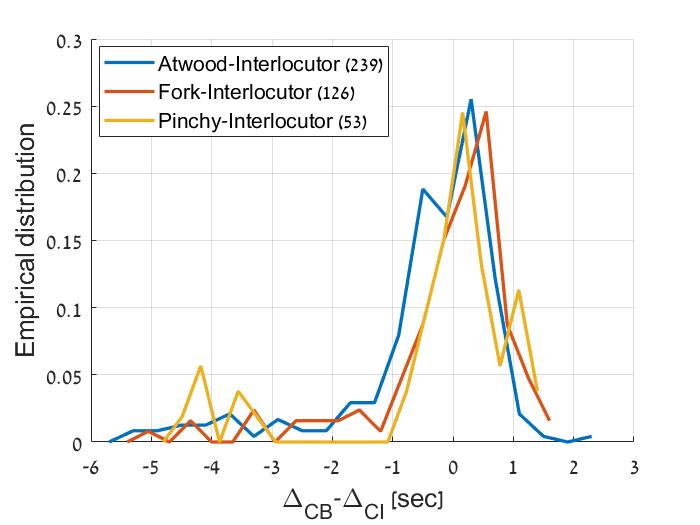}}
	\hspace{2mm}
 	\subfloat[Probability density of $\Delta_{\mathrm{CB}}$ as a function of coda type.
        \label{QA per type}]{\includegraphics[width=0.48\columnwidth]{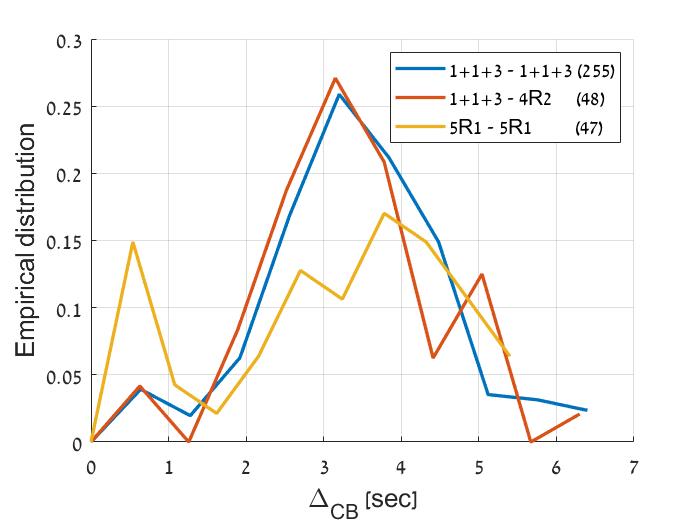}}
    \caption{Distribution functions for the delay between the end of the 'S'-coda and the beginning of the 'R'-coda, $\Delta_{\mathrm{CB}}$.}
    \label{sync}
\end{figure}

The changes in the structure of the identified codas arguably resembles that of encoding of information via a modulation signal. Here we use the analogy of a baud rate to correspond to the inter-coda Break, $\Delta_{\mathrm{CB}}$, as defined in~\ref{sec: definitions}.%, which is defined as the time interval between the reception of the last click in the 'S' coda and the first click in the 'R' coda. In this representation, possible biases due to variations in coda duration are factored out, while coda overlaps are managed by considering the 'meter' of the whales,$\Delta_{M}$.
The probability density function (PDF) of $\Delta_{\mathrm{CB}}$ is shown in Fig.~\ref{sync} for three focal whales named 'Atwood' (whale \#5586), 'Fork' (whale \#5151) and 'Pinchy' (whale \#5560). Fig.~\ref{QA per whale} shows the PDF of $\Delta_{\mathrm{CB}}-\Delta_{\mathrm{CI}}$ separately for each whale. An observable shift in the expectation value of the three PDFs reflects differences in the duration of S\&R sessions between whales. The longer tail towards the negative side of the distribution shows that sometimes whales skip a coda overlap and leave a gap of approximately similar to the exchanges Inter-coda Interval. This is further explored in Fig.~\ref{QA per type}, where we show the PDF of $\Delta_{\mathrm{CB}}$ for the three most frequent S\&R coda types, combined for the three focal whales. The seemingly similar PDFs in the latter figure show that the observed bias in Fig.~\ref{QA per whale} is not affected by the coda type.

Taken together, these results suggest that the structure of the exchange in terms of the Inter-coda Interval can be maintained despite the variation in ICI within codas, or coda types being exchanged.

%****Shane: possible meaning of results  

%, namely, $\mathrm{P}(\Delta_{\mathrm{S}}|w),~\forall T,y$. Figs.~\ref{QA per whale} and ~\ref{QA per whale} show the probability function of time lags between Q\&A coda exchanges of a given focal whale and his interlocutor, namely, $\mathrm{P}(\Delta_{\mathrm{Q\&A}}|w),~\forall T,y$, and of a given coda type, i.e., $\mathrm{P}(\Delta_{\mathrm{Q\&A}}|T),~\forall w,y$, respectively.

% 1. Fig~\ref{time lag per whale}:  The time lag between consecutive codas of a tagged whale may vary between different whales (Atwood and Fork), and thus ????
% 2. Fig~\ref{sync}: Time lag distributions between Q and A depends on the speakers and on the usage of the coda types, and thus ????
% 3. Figs.~\ref{sync} and~\ref{time lag per whale}: synchronization of time lags of consecutive codas of the tagged whale and his interlocutor (b), and thus ???? 
%synchronization of codas ICI variations of the tagged whale and his interlocutor (a). That is, both speakers use same amount of information in a conversation, and thus ???

\iffalse
\begin{figure}[]
	\centerline{\includegraphics[width=\textwidth]{Figures/Delta_Comparison_all.png}}
	\caption{Time lag dependencies}
	\label{Time lag dependencies}
\end{figure}
\fi

\subsubsection{Capacity Outage: Variation in Information Gain Encoded in Coda Types}

The analogy with the characteristics of telecommunication performance is further explored by the concept of capacity outage, which is represented here by the change in the structure of the codas, which arguably corresponds to symbol modulation and thus to information gain. The change in the coda's structure is analyzed by variations in the coda ICI relative to the typical ICI pattern for the specific coda type. This variation is measured by the metric
\begin{equation}
	\begin{aligned}
		\Delta_{\mathrm{ICI}}=||\frac{1}{L-1}\left(\sum\limits_{l=2}^{L}\tau_l(t)-\bar{\tau}_l(t)\right)||_2\;,
		\label{eq: ICI var}
	\end{aligned}
\end{equation} 
where $\tau_l(t)$ and $\bar{\tau}_l(t)$ are the ICI delays between the $l$th and $l-1$th coda's click of the measured and typical coda type $t$, $L$ is the number of clicks comprising the coda, $T$ is the number of coda types, and $||\cdot||_2$ denotes the $l2$-norm operator. %Another aspect of communication capacity is the \textit{Meter}, $\Delta_{\mathrm{M}}$, which is the time leg that passes between the emission of consecutive codas of the same whale. %Different than $\Delta_{\mathrm{S\&R}}$, metric $\Delta_{\mathrm{M}}$ considers the duration of the Coda and is thus sensitive to the Coda's type.

\begin{figure}
	\centerline{\includegraphics[width=0.6\textwidth,angle=0]{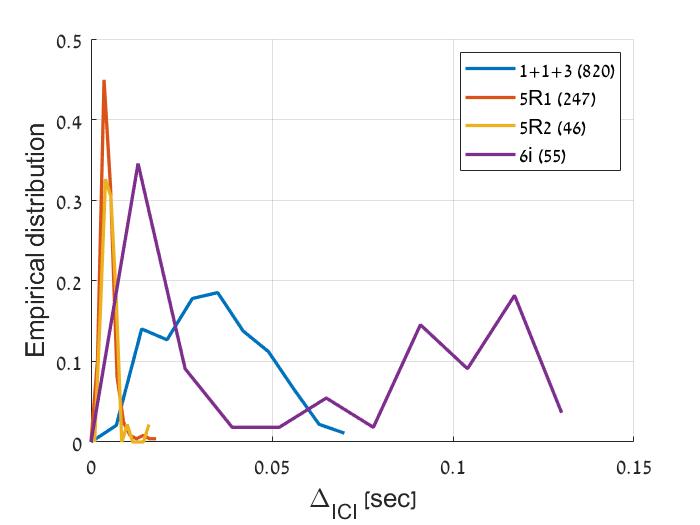}}
	\caption{PDF of $\Delta_{\mathrm{ICI}}$ as a function of coda type. The number of detected Codas per type is given in the legend in brackets.}
	\label{fig:ICI_word}
\end{figure}
The PDF of $\Delta_{\mathrm{ICI}}$ for four common S\&R coda types is examined in Fig.~\ref{fig:ICI_word}. The values of $\Delta_{\mathrm{ICI}}$ for the coda types 5R1 and 5R2 show small fluctuations which, according to our communication analogy, reflect a low degree of modulation leading to a low level of information transfer. In contrast, the PDF of $\Delta_{\mathrm{ICI}}$ for coda type 1+1+3 shows higher ICI variation with greater diversity, while the highest diversity is shown in $\Delta_{\mathrm{ICI}}$ for coda type 6i. This can be explained by the clicks structure of these coda types, where a greater ICI variation is observed the more complex the coda type is. That is, assuming that some information is encoded in the ICI variation (e.g. social identity information, ~\cite{hersh2022evidence}), the information capacity depends on the coda type. In other words, the more complex the modulation symbol (coda type) is, the more information (ICI variation) it can transport, to use the analogy with telecommunication once again.

\begin{figure}[t]
	\centering
	\subfloat[PDF of $\Delta_{\mathrm{ICI}}$ for the focal and non focal whales.
        \label{fig:ICI_variation}]{\includegraphics[width=0.48\columnwidth]{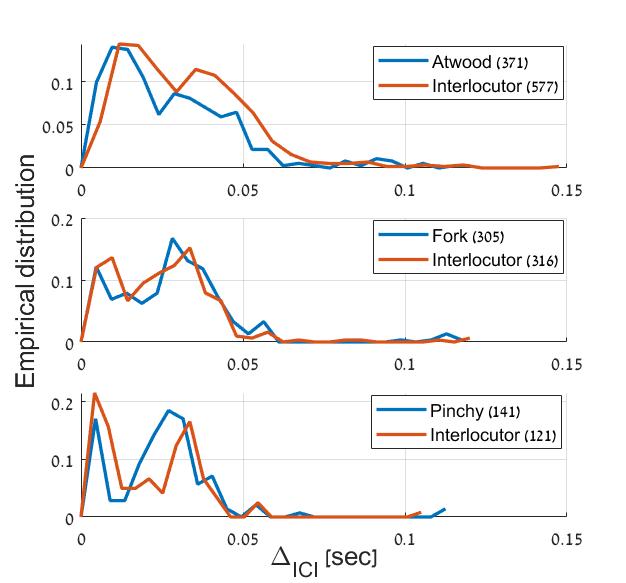}}
	\hspace{2mm}
 	\subfloat[PDF of $\Delta_{\mathrm{CI}}$ for the focal and non focal whales.
        \label{fig:time_lag}]{\includegraphics[width=0.48\columnwidth]{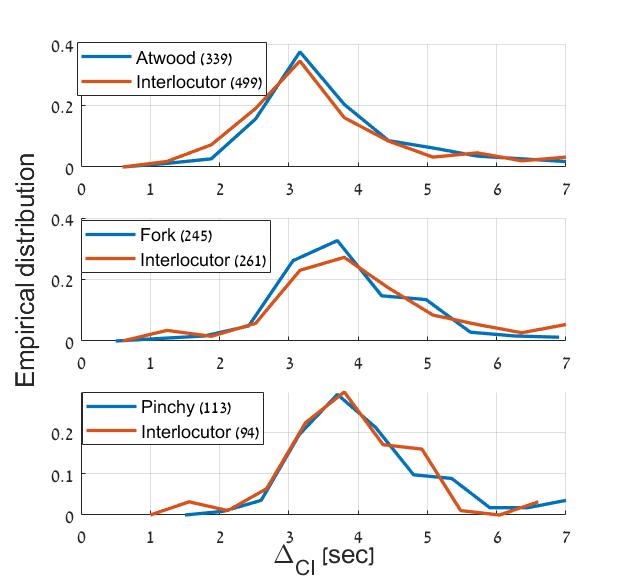}}
     \caption{Differences between variations of Codas generated by the focal and non focal whales}
    \label{f:sync2}
\end{figure} 

The above observation is next investigated for various focal whales. The PDFs of $\Delta_{\mathrm{ICI}}$ for the focal whales 'Atwood', 'Fork' and 'Pinchy' and their corresponding interlocutor are shown in Fig.~\ref{fig:ICI_variation}. The similar PDFs for the ICI of the focal and interlocutor whales reflect synchronization between the 'S'-coda and the 'R'-coda. Our results here, and in Fig.~\ref{Time lag dependencies}, suggest that whales often match coda type (and therefore $\Delta_{\mathrm{ICI}}$) during dyadic coda exchanges (as suggested by \cite{schulz2008overlapping, sharma2023contextual}. In fact, Sharma et al. \cite{sharma2023contextual}, suggests that whales in dyadic coda exchanges, like those analysed here, are able to precisely match variation in both rhythm (relative timing of clicks in a coda) and tempo (overall duration), both which drive variation in ICI, and further can do so by matching tempo (total duration of a coda) even when not matching the same coda type (rhythm). Here, we suggest that this synchronization of ICI variation may also reflect the dependence  on the identity of the emitting whale. As is supported by the differences between the PDFs for 'Atwood' (upper panel), 'Fork' (middle panel) and 'Pinchy' (lower panel) and their interlocutor are matched to the identity of the whale.

The above interpretation is supported by the results in Fig.~\ref{fig:time_lag}, where the PDF of the Inter-coda Interval, $\Delta_{\mathrm{CI}}$, is compared for the focal and the interlocutor, again for the whales 'Atwood', 'Fork' and 'Pinchy'. Again, a synchronization of $\Delta_{\mathrm{CI}}$ is observed between the signaller and the responder, a synchronization that changes depending on the identity of the different focal whales. This suggests that the whales are able to vary their Inter-coda Interval depending on who they are interacting with.

%****Shane: possible meaning of results  

% \begin{figure}[]
%     \centering
%     \begin{subfigure}[]{0.45\textwidth}
%         \centering
%         \includegraphics[height=1.7in]{Figures/ICI_var_speaker.png}
%         \caption{between whales}
%     \end{subfigure}%
%     \hfill
%     \begin{subfigure}[]{0.45\textwidth}
%         \centering
%         \includegraphics[height=1.7in]{Figures/ICI_var_new.png}
%         \caption{between coda types}
%     \end{subfigure}

%     \caption{}
%     \label{Information capacity}
% \end{figure}

\subsection{Identification of new coda types}

\begin{figure}[]
	\centerline{\includegraphics[width=140mm,angle=0]{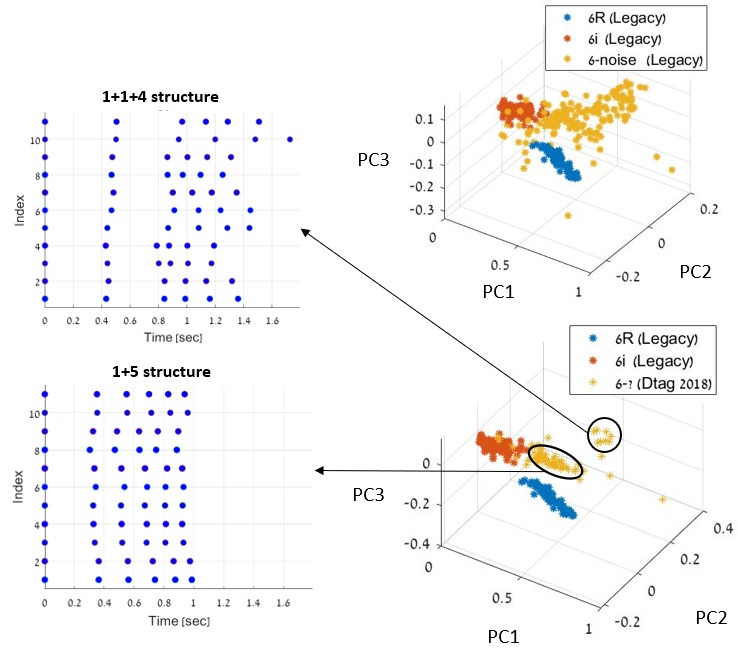}}
	\caption{Right panels: Distribution of the three dominant PCA components (PC1, PC2 and PC3) of 6 click codas out of 427 from our legacy database. The upper right panel shows the distribution of the annotated database. The bottom right panel shows the distribution of the unannotated dataset obtained with our automatic annotator and verified manually. Left panels: rhythm plots of ICIs from a few examples of the two new coda types.}
	\label{New codas}
\end{figure}

Coda types are a categorical representation of continuous ICI multivariate space. Previously, discrete coda types have been defined based on various clustering methods (see \cite{rendell2003comparing, rendell2003vocal, schulz2011individual, gero2016individual, hersh2021method}). During these processes, some of the codas, whose ICI series are outliers or located in sparse areas between dense clusters, to be labelled as 'contamination' or 'noise' (rather than being 'forced' into categories) and thus only define more conservative, dense clusters in which all codas in a discrete 'type' are highly similar to each other. An example of this is shown in Fig.~\ref{New codas}, where we visualize the three most dominant principal components, PC1, PC2 and PC3, for 427 codas with 6 clicks obtained by performing a principal component analysis (PCA) on our data. This PCA analysis allows the identification of clusters in the dataset; corresponding to different coda types. In the top right panel of Fig.~\ref{New codas} we show the analysis for the preexisting annotated dataset. We observe two prominent clusters labeled '6i' and '6R', while other scattered points are labeled as 'noise' simply because they did not meet thresholds of density or cluster shape based on the quantitative methods applied to define categories. By using our automatic detector for the near-field recordings from 2018, which were not previously annotated due to the large amount of data and the time required by manual methods, 843 codas were rapidly annotated. This then allows us to observe two more distinct clusters in the feature space, apart from the previous 6R and 6i coda types. This can be seen in the bottom right panel of Fig.~\ref{New codas}. We define two more descending class, coda types as the new types '1+5' and '1+1+4' (rhythm plots are  shown in the upper left and lower left panels of Fig.~\ref{New codas}). We note that the codas within each of the new clusters have highly similar rhythm structure with some within type tempo variation; but that the two types have distinct rhythms and tempos.

%Using the Coda detector, two new coda types were discovered. The technique for the discovery of new coda types involves clustering the codas by features. A Coda type is defined by the signal's preamble sequence of 1-3 equally spaced clicks, followed by an encoding part of 2-10 clicks with shorter spacing. While the total number of clicks within the Coda defines its type, the Coda's style is determined by the spacing between the clicks in the encoding part. Three Coda classes are documented: 1) 'R': regular (or equal) spacing, 2) 'i': increasing spacing, and 3) 'D': decreasing spacing. To find new coda types, projection to two key features is performed to the ICI sequence of each detected coda. The two features are determined by the results of a principle component analysis (PCA) over the codas' ICI sequences. to discover new coda types, a comparison is made between the 2D distribution of the ICI's two key features of the explored dataset and of historical data (DSWP: 2005-2016, ****Guy(give reference)).

%The results in Fig.~\ref{New codas} for the 2D distribution of key features for codas comprising 6 clicks shows two new prominent clusters. These Codas have distinguished types 1+1+4, which corresponds to two component preamble consists of two long equal spacing between the first three clicks followed by 4-component encoding, and 1+5, which reflects one component preamble of one long equal spacing between the initial click pair followed by 5-component encoding. To the best of our knowledge, these two types are not described in the literature. 

\section{Discussion}

Our approach, arguably the first automatic detector and annotator for sperm whale codas, offers robust operation with few parameters to set, allowing analysis of data obtained by both acoustic tags and boat-mounted hydrophones located far from the whale. The ROC performance of the detector shows that it is robust even at low SNR values and in high dynamic range, that it can handle the noise transients commonly found in the Dominica Island, and that it is able to distinguish between echolocation clicks and coda. As shown in Section~\ref{sec:method} below, our annotator can also separate between overlapping codas. When the analyzed buffer is small, there are limitations when echolocation clicks from different whales overlap, as they may resemble a coda structure. Future developments will therefore focus on the inclusion of blind source separation to mitigate cases of interfering clicks, as well as a temporal likelihood prediction model to allow generalization to a larger number of coda types.

The detector provides automatic analysis of large datasets of raw acoustic data, thereby significantly reducing the processing time. The results of the automatic annotation were demonstrated to explore the characteristics of the codas. In particular, the distribution of coda types, the delays between successive codas and the ICI variation for different coda types and between different focal and non-focal whale pairs. An analogy was made for the resemblance of coda exchange to a telecommunications session, where the synchronization between the focal whale and its interlocutor in terms of delays between successive codas was compared with the baud rate of the communication, and the synchronization in terms of ICI variation was attributed to capacity outage. The large amount of codas collected with the new detector allowed a statistical analysis of coda features to comment on the diversity of codas. This led to the discovery of two new coda types.

Significantly, our results suggest that some of the within coda type variation, the structure of coda exchanges in terms of the inter-coda-intervals and inter-coda-breaks, vary based on the identity of interlocutors. While \cite{gero2016individual} suggested that within coda type ICI variation of the 5R1 coda could be used to potentially recognize individuals within a small number of social units, these new findings suggest identity cues at the level of the structure of the coda exchange. This will require deeper investigation in relation to if these patterns are upheld depending on which whale vocalizes first (vocal asymmetries), such that whale B may match whale A's inter-coda-interval or inter-coda breaks when A vocalizes first, but A will match B if B vocalizes first. But also, how these patterns may perhaps reflect vocal cues of social dominance (see ~\cite{cheng2016listen, kavanagh2021dominance} or correlate with social or kin based relationships. Interestingly, Fork and Pinchy belong to the same social unit, while Atwood lives in a different one; and their distributions of $\Delta_{\mathrm{CI}}$ are more similar to one another than to Atwood's distribution; so these patterns may reflect unit-level, rather than individual level patterns. %While our current dataset is not able to distinguish between these hypotheses, the creation and validation of this automated annotator can quickly process the back catalogue of recordings of identified whales collected at our field site since 2005.

Looking towards the future plans for the detector, we intend a real-time implementation within the framework of Project CETI. This will include online detection from an offshore moorings to support the tagging work and create a live map of the whales' positions. The detector will also be implemented in real-time on board a sea glider and in surface vessels searching the area for sperm whales.

\section{Methods}\label{sec:method}

\subsection{Method of Data Collection}

Data collection was conducted in two ways: 1) acoustic tags attached to sperm whales (near-field), and 2) acoustic recorders deployed from a boat (far-field). In the first method, sound and movement tags (Dtag generation 3 \cite{DtagPaper}) worn by the animals were deployed between 2014 and 2018 as a part of The Dominica Sperm Whale Project. These tags have a two-channel audio sampling at 120~kHz with a resolution of 16~bit, providing a flat ($\pm$2~dB) frequency response between 0.4~kHz and 45~kHz. Pressure and acceleration were also sampled at a rate of 500~Hz and a resolution of 16~bit, decimated to 25~Hz for offline analysis. Tagging was carried out from an 11~m rigid-hulled inflatable boat (RHIB). A 9~m hand-held carbon fiber pole was used to attach individual tags to a whale with four suction cups.

Data collection in the far field was carried out with a C75 Cetacean Research hydrophone deployed from the RIB at a depth of 15~m. The hydrophone has a flat ($\pm$3~dB) frequency response between 10~Hz and 170~kHz. A MicPre-6 || data acquisition device from 'Sound Devices' was used to sample the signal at 96~kHz and 16-bit resolution. 

All whales were identified from photographs of the trailing edge of their fluke ( \cite{arnbom1987}). The identifications were used to ensure that only recordings from one of the two sympatric clans (EC1 – the Eastern Caribbean Clan) were included in the analysis to control for any differences in repertoire between vocal clans \cite{Gero2016-social}. 

\subsection{Permitting and Animal Care Approval}
Data from The Dominica Sperm Whale Project were collected under scientific research permits from the Fisheries Division of the Government of Dominica. The field protocols for approaching, photographing, non-invasively tagging, and recording sperm whales were approved by either the University Committee on Laboratory Animals of Dalhousie University, Canada; the Animal Welfare and Ethics Committee of the University of St Andrews, Scotland; or Aarhus University, Denmark; and sometimes several or all of these across years. Data collected as a part of Project CETI were also collected under a scientific research permit from the Fisheries Division of the Government of Dominica. The field protocols for recording sperm whales for Project CETI were approved by The Institutional Animal Care \& Use Committee at Harvard University. All methods are reported in accordance with ARRIVE guidelines relevant to wild whales and our analyses.
  
\subsection{Coda Detection}

\begin{figure}[]
	\centerline{\includegraphics[width=135mm,angle=0]{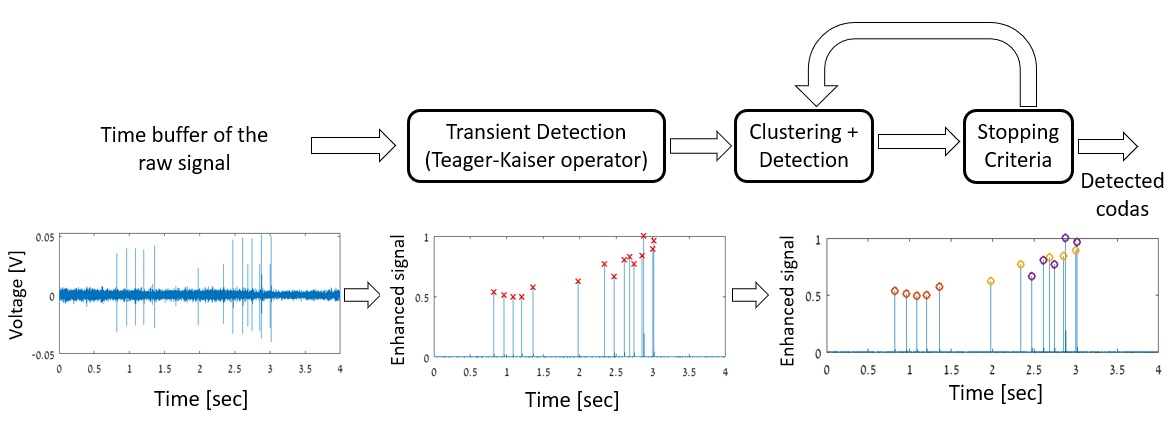}}
	\caption{A block diagram for the Coda detector.}
	\label{Fig.1: Detection block diagram}
\end{figure}
A block diagram of the coda detector and annotator is shown in Fig.~\ref{Fig.1: Detection block diagram}. The input to the detector is a buffer of $7~\mathrm{sec}$ (configurable). First, a transient detector identifies a few regions- of interest (ROIs). For each ROI, three features are extracted: 1) the shape of the transient, 2) the intensity of the transient, and 3) a measure of the multipulse structure, if it exists. Clustering is then performed based on these features to group the ROIs based on their likelihood similarity. The clustering is performed iteratively until a stopping criteria is met. The details of the individual components are described below. A MATLAB implementation code for our scheme is available in \cite{Webside_coda}.

\subsubsection{Click Identification}

Transient ROIs are determined by selecting identified peaks based on the expected attributes of valid coda clicks. The process starts with a bandpass filter for the desired frequency range of 2~kHz-24~kHz \cite{lohrasbipeydeh2014adaptive}. To enhance transients, following prior works for sperm whale click detection~\cite{gubnitsky2023inter,lohrasbipeydeh2014adaptive}, the Teager-Kaiser energy operator (TKEO) ~\cite{kaiser1990simple},
\begin{equation}
	\begin{aligned}
		z_n= x_n^2-x_{n-1} \cdot x_{n+1}\quad, n=1,\ldots,N \;.
		\label{eq: TKEO}
	\end{aligned}
\end{equation}
is executed over the sampled signal $x_n, / n=0,\ldots,N-1$. Peaks in $z_n$ from \eqref{eq: TKEO} are identified as local energy maxima that fulfill a minimum distance between the peaks and a minimum threshold for the SNR. The former is set according to the maximum IPI of known codas (a value of 8~ms is used in the results). We limit the number of identified transients to an expected maximum of 20 peaks. An example of the transient identification process is shown in Fig.~\ref{Fig.1: Detection block diagram}.

\iffalse
the peak's surrounding. Specifically, assuming non-transient samples $z_n$ in \eqref{eq 2} follow the Rayleigh distribution \cite{bergmann1946intensity}, 
\begin{equation}
	\begin{aligned}
		R_{(z_n)}= \frac{z_n}{b^2}\exp^{\left( \frac{-z_n^2}{2b^2} \right)} \;,
		\label{eq Rayleigh distribution}
	\end{aligned}
\end{equation} 
where $b$ is a scaling parameter, the value threshold is set by maximum likelihood. Expecting a small number of transients in the buffer $\bf{z}=\left[z_1,\ldots,z_N\right]$, a maximum likelihood estimation for $b$ is \cite{siddiqui1964statistical} 
\begin{equation}
	\begin{aligned}
		\hat{b}= \sqrt{\frac{1}{2N}\bf{z}^T\bf{z}} \;.
		\label{eq Rayleigh distribution}
	\end{aligned}  
\end{equation} 
Finally, the transient's value threshold, $\rho_v$, is set by the constant false alarm test,
\begin{equation}
\mathrm{FAR}= \int_{\infty}^{\rho_v}R_{(z_n)}dz_n\;,
\label{eq:threshol_val}
\end{equation} and, 
\fi
%Expecting a coda to include not more than 20 clicks,

\subsubsection{Clustering transients into candidate codas}
\label{sec: Clustering}

The clustering of transient ROIs into groups of possible codas is based on the observation that codas follow known ICI patterns~\cite{gero2016individual} and on the assumption that the clicks forming a coda share statistical similarities. There are two explanations for the latter assumption: 1) each of the coda clicks is generated by the emitting whale in the same way, and 2) due to the short duration of the coda ($0.1-2sec$)~\cite{pavan2000time,amano2014differences,gero2016individual}), the channel's impulse response from the whale to the receiver is expected to be roughly time invariant, and thus the clicks of the same coda share the same channel distortion. An example of this is shown in Fig.~\ref{Challenges}, where different spectral shapes are observed for the far-field and near-field codas.

%Conversely, the CIR may change significantly with time as the whale changes its orientation and distance with respect to the receiver, hence compromising the similarity between the clicks' structure. An example is illustrated in Fig.~\ref{Challenges} where two coda spectrograms of a focal and a distant non-focal whales are shown. Although the clicks' tempo-spectral structure of the distant coda changes significantly, the coda's components remain similar to each other.

The clustering is formalized as the following optimization problem 
\begin{subequations}\label{complex optimization problem}
\begin{align}
\hat{K}, \hat{\bf{C}} =&\mathrm{argmax}_{K, \bf{c}_k, \ k=1,\ldots,K}\frac{1}{K}+\sum_{k=1}^{K}{\cal J}(K, \bf{c}_k)_{k=1,\ldots,K}& \label{e:Utility}\\
&\;\mathrm{s.t.}\;\;\mathcal{J}_{(\mathbf{c}_k)}>\rho_d
\label{e:Const1}\\
&\;\qquad c_l \perp  c_k,~\forall l \neq k\;,\label{e:Const2}\\
&\;\qquad \mathcal{P}(\mathbf{c}_k)>\mathcal{P}_{\mathrm{min}}\;,\label{e:Const3}\\
&\;\qquad f_r(\mathbf{c}_k)<f_r^{\mathrm{max}}\label{e:Const4}\;.
\end{align}
\end{subequations}
where $\mathbf{c}_k, \ k=1\ldots_K$ are M-dimensional binary assignment vectors such that 
\begin{equation}
\mathbf{c}_k=[\omega_{k,1},\ldots,\omega_{k,M}]\;, \quad \omega_{k,l}\in\{0,1\}, \ l=1,\ldots,M\;.
\label{eq1}
\end{equation}
The utility function $\mathcal{J}_{(\mathbf{c}_k)}$ is defined by
\begin{equation}
\mathcal{J}_{(\mathbf{c}_k)}= \mathcal{L}^\mathrm{s}(\mathbf{c}_k)+\alpha_1 \mathcal{L}^\mathrm{t}(\mathbf{c}_k)-\alpha_2 F(\mathbf{c}_k)\;,
\label{eq1_1}
\end{equation}
where $\mathcal{L}^\mathrm{s}(\mathbf{c}_k)$ and $\mathcal{L}^\mathrm{t}(\mathbf{c}_k)$ are the structural and temporal similarity likelihoods for a clustering vector $\mathbf{c}_k$, respectively, $F(\mathbf{c}_k)$ is a penalty factor for low rank clusters such that
\begin{equation}
F(\mathbf{c}_k)=\exp{({\mathbf{c}_k \mathbf{c}_k^\mathrm{T}})^{-1}}\;,
\label{eq1_2}
\end{equation}
and $\alpha_1$ and $\alpha_2$ are normalized weighting factors. Note that $\alpha_1$ is a limiting factor intended to restrict the selection of clusters that define 'sub-codas'. For the case shown in Fig.~\ref{Fig.1: Detection block diagram} for example, a large $\alpha_1$ will favor the choice of the entire structure 4+1 over the sub-structure of 4 clicks with identical ICI, although both represent valid coda structures. On the other hand, a small $\alpha_1$ allows the identification of new, unknown or rare coda types.

Four constraints are used. In \eqref{e:Const1} a utility function above a detection threshold $\rho_d$ is searched for. In \eqref{e:Const2} we restrict the solution to a unique clustering. In \eqref{e:Const3}, we expect a valid coda click to contain at least $\mathcal{P}_{\mathrm{min}}$ multipulses and limit the number of multipulses in a click, averaged over $\mathbf{c}_k$, to $\mathcal{P}(\mathbf{c}_k)$. While both coda and echolocation clicks consist of repetitive pulses of decreasing intensity, codas are characterized by a much lower decay rate~\cite{madsen2002sperm}. This is possible thanks to the whale's ability to use the right nasal passage as an adaptive acoustic valve to direct the signal towards the junk in the case of echolocation clicks, or back in the spermaceti in the case of coda signals~\cite{huggenberger2014acoustic}. The low decay rate of the intermediate pulses is reflected in a larger number of pulse repetitions, resulting in a more pronounced periodic structure of the overall signal. A method for estimating the number of multipulses is presented in Section~\ref{Interpulse analysis}. Our fourth condition in \eqref{e:Const4} considers that the resonant frequency of the coda click, averaged over $\mathbf{c}_k$, $f_r(\mathbf{c}_k)$, should be below a threshold $f_r^{\mathrm{max}}$, and following \cite{madsen2002sperm}, we use $f_r^{\mathrm{max}}=12$~kHz. The resonant frequency is calculated using a super-resolution spectrogram in~\cite{moca2021time}.

Due to the requirement for unique clustering, the optimal solution of \eqref{complex optimization problem} is an NP-hard problem with similarities to the graph coloring problem~\cite{garey1997computers}. In the case of many ROI candidates, a suboptimal solution is offered based on the following iterative procedure:
\begin{subequations}\label{e:iterative}
\begin{align}
\hat{c}_k =&\mathrm{argmax}_{c_k}\mathcal{J}_{(\mathbf{c}_k)}& \label{e:IterativeUtility}\\
&\;\mathrm{s.t.}\;\;\mathcal{J}_{(\mathbf{c}_k)}>\rho_d
\label{e:iterativeConst1}\\
&\;\qquad \mathcal{P}(\mathbf{c}_k)>\mathcal{P}_{\mathrm{min}}\;,\label{e:iterativeConst2}\\
&\;\qquad f_r(\mathbf{c}_k)<f_r^{\mathrm{max}}\label{e:iterativeConst3}\;.
\end{align}
\end{subequations}
In each iteration, cluster candidates $\mathbf{c}_k$ that do not meet the orthogonality condition,
\begin{equation}
\begin{array}{rrclcl}
 c_l \cdot \hat{c}_k^T=0,~\forall l\neq k.
\end{array}
\end{equation} are omitted.
The iterations stop when no more valid clusters are found. 

% Although solving \eqref{complex optimization problem} may lead to the optimal solution, it requires an exhaustive search. In fact, Eq.~\eqref{complex optimization problem} is an NP hard optimization problem known as the graph coloring problem~\cite{garey1997computers}. A suboptimal but efficient solution is possible by iteratively identifying a single valid cluster. The optimization problem becomes:
% \begin{equation}
% \begin{array}{rrclcl}
% \hat{c}_k=\displaystyle \max_{c_k} & \multicolumn{3}{l}{\mathcal{J}_{(\mathbf{c}_k)}}\\
% \textrm{s.t.} & \mathcal{J}_{(\mathbf{c}_k)}>\tau
% \end{array}
% \end{equation}
% In each iteration we eliminate transient candidates associated with the detected cluster. Formally, we eliminate all cluster candidates that do not meet the following orthogonality condition:
% \begin{equation}
% \begin{array}{rrclcl}
%  c_l \cdot \hat{c}_k^T>0,~\forall l.
% \end{array}
% \end{equation}
% The algorithm stops when no more clusters are found that exceed the threshold $\tau$.

%\begin{equation}
%\mathcal{J}_{(\mathbf{c}_k)}=\frac{\mathbf{c}_k A \mathbf{c}_k^\mathrm{T}}{0.5(\mathbf{c}_k \mathbf{c}_k^\mathrm{T}-1)\mathbf{c}_k \mathbf{c}_k^\mathrm{T}}+e^{-\alpha\mathbf{c}_k \mathbf{c}_k^\mathrm{T}}+\beta \mathcal{L}_{(\mathbf{c}_k)}^{-1}.\label{eq1}
%\end{equation}

\subsubsection{Structural likelihood analysis}

To represent the similarity between a transient pair, each detected transient is treated as a node in an undirected graph. Let $s_{i,j}$ denote the similarity between a node pair $i$ and $j$, where $s_{i,i}=0$. These similarities combine an affinity matrix 
\[
  S_{M\times M} =
  \left[ {\begin{array}{cccc}
    s_{11} & s_{12} & \cdots & s_{1M}\\
    s_{21} & s_{22} & \cdots & s_{2M}\\
    \vdots & \vdots & \ddots & \vdots\\
    s_{M1} & s_{M2} & \cdots & s_{MM}\\
  \end{array} } \right]\;.
\]
For a candidate cluster, $\mathbf{c}_k$, the structural likelihood is defined as the average similarity between the cluster's nodes
\begin{equation}
\mathcal{L}^\mathrm{s}(\mathbf{c}_k)=\frac{\mathbf{c}_k S \mathbf{c}_k^\mathrm{T}}{0.5(\mathbf{c}_k \mathbf{c}_k^\mathrm{T}-1)\mathbf{c}_k \mathbf{c}_k^\mathrm{T}}\;.
\label{eq1_3}
\end{equation} 

Three similarity measures are used: 1) waveform-based similarity, $s_{i,j}^{\mathrm{shape}}$, 2) IPI-based similarity, $s_{i,j}^{\mathrm{IPI}}$ and 3) intensity-based similarity, $s_{i,j}^{\mathrm{I}}$. The first one is calculated by the cross-correlation between the pair of clicks $y_i(t)$ and $y_j(t)$. Formally,
\begin{equation}
s_{i,j}^{\mathrm{shape}}=\frac{\int y_i(t)\cdot y_j(t)dt}{\sqrt{\int y_i(t).^2dt}\cdot \sqrt{\int y_j(t).^2dt}}\;.
\label{shape similarity}
\end{equation}
The second similarity metric requires the evaluation of the click's IPI e.g., using the method in \cite{gubnitsky2023inter}, and is defined by the normalized difference
\begin{equation}
s_{i,j}^{\mathrm{IPI}}=1-\frac{|\mathrm{IPI}_i-\mathrm{IPI}_j|}{\mathrm{max}(\mathrm{IPI}_i,\mathrm{IPI}_j)}\;.
\label{IPI-similarity}
\end{equation}
Since the IPI is a function of the spermaceti size~\cite{gordon1991evaluation}, the measure $s_{i,j}^{\mathrm{IPI}}$ allows the separation between overlapping clicks of whales of different sizes. Similarly, the intensity-based similarity is defined by the difference
\begin{equation}
s_{i,j}^{\mathrm{I}}=1-\frac{|\mathrm{I}_i-\mathrm{I}_j|}{\mathrm{max}(\mathrm{I}_i,\mathrm{I}_j)}\;,
\label{In tensity similarity}
\end{equation}
where $\mathrm{I}_i$ is the RMS amplitude of the $i$th click in the cluster. Since underwater acoustic propagation is strongly dependent on distance, $s_{i,j}^{\mathrm{I}}$ allows the discrimination of clicks emitted by whales at different distances from the receiver. The combined similarity measure $s_{i,j}$ is defined by the normalized weighted sum
\begin{equation}
s_{i,j}=\rho^{\mathrm{corr}}s_{i,j}^{\mathrm{corr}}+ \rho^{\mathrm{IPI}} s_{i,j}^{\mathrm{IPI}}+ \rho^{\mathrm{I}} s_{i,j}^{\mathrm{I}}\;,
\label{eq1_4}
\end{equation} 
where $\rho^{\mathrm{corr}}+\rho^{\mathrm{IPI}}+\rho^{\mathrm{I}}=1$.

\subsubsection{Temporal likelihood analysis}

%Let $L_W$ be the number of possible clusters, each containing exactly $W$ values of ICI from $W+1$ clicks. Also let $\mathbf{O} \in \mathrm{I\!R}^{L_W \times W}$ be a matrix whose element $o_{i,j}$ corresponds to the ICI between the $j$th and the $j+1$th clicks in the $i$th coda. Similarly, let $\mathbf{H}$ %\mathbf{H} \in \mathrm{I\!R}^{L_W \times W}$. Matrix $\mathbf{H}$ is compressed to yield matrix ${\mathbf{H}}_R$ using the principle component analysis (PCA) to find the three eigenvectors that correspond to the three largest eigenvalues of the sampled correlation matrix of ${\mathbf{H}}$.
Let $\mathbf{H}$ be a database of ICI patterns of $L_W$ previously collected codas, each including exactly $W+1$ clicks. We use $\mathbf{H}$ to evaluate the likelihood of a possible cluster $c_k$ to represent a coda signal. Features of $\mathbf{H}$ are extracted by principle component analysis (PCA) to form a $Q \times W  ~( Q < L_W)$ representative matrix
\begin{align}	\mathbf{H}_\mathrm{R}=\mathbf{\overline{\mathbf{H}}}\cdot{\overline{\mathbf{U}}}\;,
	\label{pca database}
\end{align}
 where $\overline{\mathbf{H}}$ is an unbiased version of ${\mathbf{H}}$ in which the sampled mean of each column of ${\mathbf{H}}$ is shifted to zero, and $\overline{\mathbf{U}}\in\mathrm{I\!R}^{W \times {Q}}$ is a matrix representation of the $Q$ eigenvectors corresponding to the largest $Q$ eigenvalues of ${\mathbf{H}}$. Matrix ${\overline{\mathbf{U}}}$ is used also to extract the same $Q$ features from the ICI pattern, $\mathbf{ICI}_{(\mathbf{c}_k)}$, of the examined potential $c_k$ cluster, denoted by
\begin{align}	\mathbf{O}_{(\mathbf{c}_k)}=\mathbf{ICI}_{(\mathbf{c}_k)}\cdot{\overline{\mathbf{U}}}\;.
	\label{observation cluster}
\end{align} 
To evaluate the probability of $\mathbf{O}_{(\mathbf{c}_k)}$ to be a valid coda representation, we calculate the temporal likelihood
\begin{equation}
\mathcal{L}^\mathrm{t}_{(\mathbf{c}_k)}= \sum_{g=1}^{G_W} P_r(\mathbf{O}_{(\mathbf{c}_k)}\implies g|\Theta_W)\;,
\label{e:likelihood}
\end{equation}
where the $\mathbf{O}_{(\mathbf{c}_k)}\implies g$ term implies that $\mathbf{O}_{(\mathbf{c}_k)}$ is a feature representation of coda type $g$, $G_W$ is the number of coda types available in $\mathbf{H}$, and $\Theta_W$ is the set of parameters for the distribution of the coda features in $\mathbf{H}_\mathrm{R}$. 

To calculate \eqref{e:likelihood}, we evaluate $\Theta_W$ from ${\mathbf{H}}_R$. We estimate the distribution of each coda type $g$ separately. Since the coda type is associated with the number of clicks, in the following we drop the sub-index $W$. Denote $h_g \in \mathbb{R}^{1 \times Q}$ as a subset of $\mathbf{H}_\mathrm{R}$ corresponding to coda type $g$, and let $\Theta_g$ be its distribution parameters. We model $h_g$ as the mixture model
\begin{equation}
 p(h_g|\Theta_g)= \sum_{k=1}^{K_g} \phi_k \cdot f_k(h_g | \theta_{g,k})
 \label{eq:GMM}
\end{equation}
of $K_g$ clusters with prior 
\[
\sum_{k=1}^{K_g} \phi_k = 1
\]
and the generalized Gaussian distribution\footnote{The General Gaussian distribution is chosen by its flexibility to represent multiple distribution types through its shape parameter $\beta$ \cite{diamant2012and}} \cite{kotz1975multivariate},
\begin{equation}
f_k(h_g | \theta_{g,k} ) = \frac{\Gamma(\frac{Q}{2})}{\pi^{\frac{Q}{2}}\Gamma(\frac{Q}{2\beta_{g,k}})2^{\frac{Q}{2\beta_{g,k}}}} \frac{\beta_{g,k}}{m^{\frac{Q}{2}}|\Sigma_{g,k}|^{\frac{1}{2}}} \cdot \exp\{-\frac{1}{2m^{\beta_{g,k}}} (h_g-\mathbf{\mu}_{g,k})^T \Sigma_{g,k}^{-1} (h_g-\mathbf{\mu}_{g,k})\}\;.
\label{eq:general gaussian}
\end{equation}
where $m\in \mathbb{R}^{1 \times Q}$ is a vector of scale parameters, $\theta_{g,k}=\{\mathbf{\mu}_{g,k}, \Sigma_{g,k}, \beta_{g,k}\}$ is the distribution parameter set with the expectation vector $\mathbf{\mu}_{g,k}\in \mathbb{R}^{1 \times Q}$, the shape vector $\beta_{g,k}\in \mathbb{R}^{1 \times Q}$ and the symmetric covariance matrix $\Sigma_{g,k}\in \mathbb{R}^{Q \times Q}$.

For $L_g$ being the number of codas of type $g$ in $\mathbf{H}$ and $h_g^l$ being the corresponding PCA-induced feature vector, we find the distribution parameters $\hat{\Theta}_{g} =\{\phi_k,m,\theta_{g,k}\}$ by the maximum likelihood
\begin{equation}
\begin{aligned}
\hat{\Theta}_g=\argmax_{\Theta_g} \quad &  \sum_{l=1}^{L_g} \log\left(p(h_g^l|\Theta_g)\right)\;,
\end{aligned}  
\label{e:Parameter_estimation}
\end{equation} 
To solve \eqref{e:Parameter_estimation} we use a combination of the expectation maximization (EM) approach with the Riemannian averaged fixed-point (RA-FA) learning method according to the steps presented in~\cite{boukouvalas2015new} and~\cite{najar2018unsupervised}, respectively.

Note that the above method requires the estimation of the number of clusters, $K_g$, in \eqref{eq:GMM}. To estimate $K_g$, we consider the Bayesian information criterion (BIC) \cite{schwarz1978estimating}, which is effective when the dimensionality is small and the sample size $L_g$ is larger than the distribution parameters. Applied to the above case 
\begin{equation}
   \hat{K}_g=\underset{K_g}{\mathrm{argmin}}(\mathrm{BIC})\;,
   \label{e:bic}
\end{equation} 
where 
\begin{equation}
\mathrm{BIC}=\mathrm{ln}(L_g)\left(\sum\limits_{k=1}^{K_g}(|\theta_{g,k}|)-1\right)-2\cdot \mathrm{ln}\left(\sum_{l=1}^{L_g} \log\left( p(h_g^l|\hat{\Theta}_g)\right)\right)\;.
\end{equation} 
Finally, plugging \eqref{eq:GMM} and \eqref{e:Parameter_estimation} into \eqref{e:likelihood}, the temporal likelihood is calculated by
\begin{equation}
\mathcal{L}^\mathrm{t}_{(\mathbf{c}_k)}= \sum_{g=1}^{G_W} p(\mathbf{O}_{(\mathbf{c}_k)}|\hat{\Theta}_g)\;.
\label{eq:temporal likelihood}
\end{equation}

%is the $q$ th component density, $\Gamma(\cdot)$ is the Gamma function and $\Theta$ is the model parameter matrix whose vector $\theta_{g,q}=\{\mathbf{\mu_{g,q}}, \mathbf{\Sigma_{g,q}}, \mathbf{\beta_{g,q}}\}$ denote the mean, scale matrix, and shape parameter, respectively, of the $q$ th component in word $g$. And the parameters $\phi_q$ are the mixing weights such that $\phi_q > 0 ~ (q = 1 , ..., Q_g)$ and  $\sum_{g=1}^{G_Q} \phi_g = 1$.
%The parameter $\beta$ controls the peakedness of the distribution and the heaviness of its tails. %When $\beta=1$ we retrieve the multivariate Gaussian (MG) distribution, in this case $\Sigma$ becomes the covariance matrix. When $\beta=\frac{1}{2}$ , we obtain the multivariate Laplacian (ML) distribution. Furthermore, when $\beta \to \infty$  the MGG distribution becomes a uniform distribution. If the assignment of each data-point to its cluster is known a-priori we can estimate separately for each cluster the MGG's parameters, then normalize and sum them together. To that end, we use the Riemannian averaged fixed-point (RA-FP) learning approach [] as it showed accurate estimations for $\beta >1$ values. An overview of the model's parameter estimation procedure is presented in the Appendix 1.

\begin{figure}[]
    \centering
        \centering
        \includegraphics[height=2in]{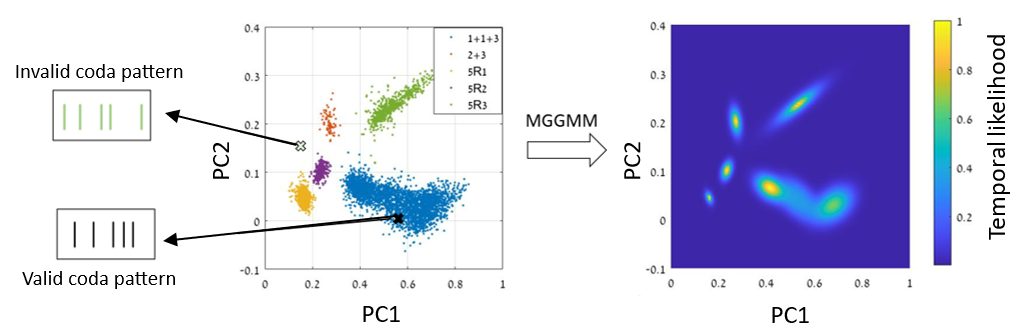}
\caption{Left: distribution function from our database for the two significant eigenvectors, PC1 and PC2, for codas with 5 clicks. Right: projection map representing the temporal likelihood that an observed cluster of transients is associated with a valid coda.}
\label{Likelihood analysis}
\end{figure}
Fig.~\ref{Likelihood analysis} shows a distribution function from our database for codas of 5 clicks and its projection into the temporal likelihood by the above procedure. Each point in the projection map represents the probability that a potential cluster of ICIs represented by the principal component represents a valid coda. We observe five non-overlapping clusters, each representing a different type of coda: 2+3, 5R1, 5R2, 5R3 and 1+1+3. By \eqref{e:bic}, the first four codas are represented by $\hat{K}_g=1$, while for the 1+1+3 coda type we estimated $\hat{K}_g=3$.

\subsubsection{Interpulse analysis}
\label{Interpulse analysis}
Next, we discuss the process of calculating the average number of multipulses $\mathcal{P}(\mathbf{c}_k)$ for \eqref{e:Const3} to separate codas from echolocation clicks. To identify interpulses, we use phase-slope analysis. Based on the transient nature of pulses, an interpulse is determined at a positive zero crossing of the phase slope function (PSF)~\cite{kandia2008phase}
\begin{equation}
	\begin{aligned}
		\mathrm{PSF}_n=-\frac{1}{J} \sum_{\omega_1}^{\omega_J } \frac{X_R(\omega) Y_R(\omega)+X_I(\omega) Y_I(\omega)}{X_R^2(\omega) + Y_R^2(\omega)}\;,
	\end{aligned}
 \label{e:PSF}
\end{equation} 
where $X(\omega)=X_R(\omega) + jX_I(\omega),$ and $~~ Y(\omega)=Y_R(\omega) + jY_I(\omega)\;,$ are the Fourier transforms of $z_n$ from \eqref{eq: TKEO} and $n z_n$, respectively, and $R$ and $I$ stand for the real and imaginary parts, respectively. 

\begin{figure}[t]
	\centering
	\subfloat[\label{fig: psf coda tag}]{\includegraphics[width=0.48\columnwidth]{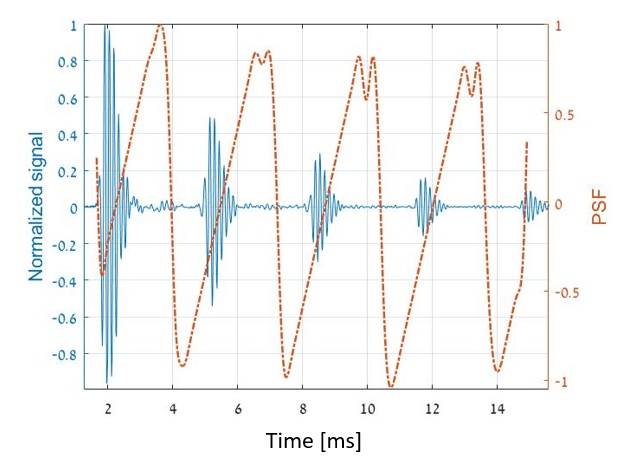}}
	\hspace{2mm}
 	\subfloat[\label{fig: psf coda remote}]{\includegraphics[width=0.48\columnwidth]{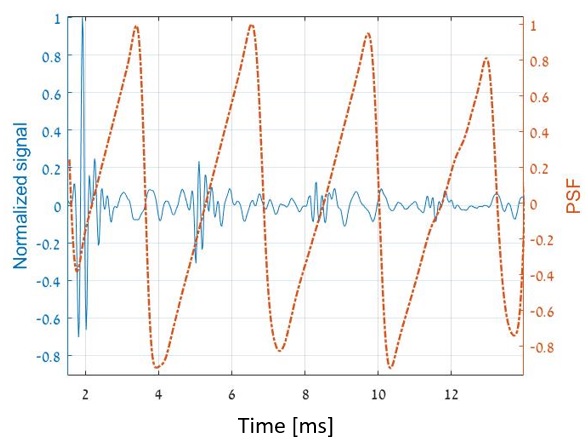}}
  	\hspace{2mm}
 	\subfloat[\label{fig: psf echolocation remote}]{\includegraphics[width=0.48\columnwidth]{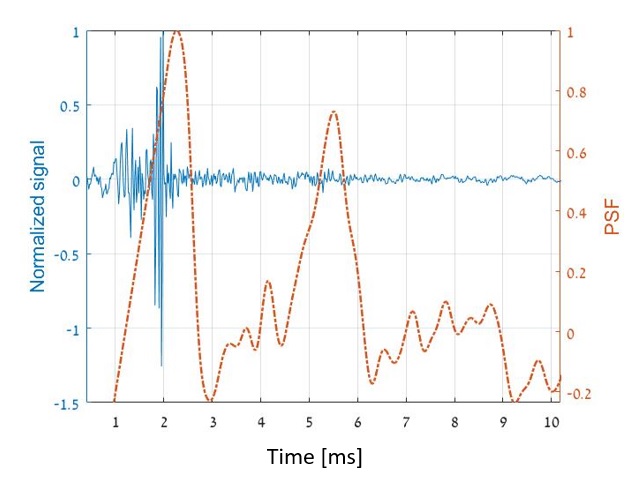}}
    \caption{(a) Tag measurement of a coda click, (b) Remote measurement of a coda click, (a) Remote measurement of echolocation click. The corresponding estimated PSF from \eqref{e:PSF} is marked in dashed red lines. Each positive zero crossing of the PSF indicate the presence of a pulse.}
    \label{fig: PSF}
\end{figure}

The number of identified interpulses $\mathcal{P}(\mathbf{c}_k)$, is calculated by the number of positive zero crossings in the PSF across the ROI of the cluster $c_k$. Fig.~\ref{fig: PSF} shows an example of the calculated PSF of code clicks with high and low SNR as well as of echolocation click. The examples illustrate the robustness of the method to weak signals, with more interpulses observed for coda clicks even at low SNR. We note that more pulses are expected for an echolocation click in the case of on-axis, i.e., when the whale's head is pointed at the receiver. However, this is a rare scenario.

\subsection{Examples of Detection and Annotation}

\begin{figure}[t]
	\centering
	\subfloat[\textbf{\textit{Wide dynamic range}}: example of near-field annotation for two codas, one with high SNR from the focal whale (red markers) and one with low SNR from a distant non-focal whale (yellow markers). A zoom in of the low SNR coda is shown in the top left box.
        \label{fig:Example1}]{\includegraphics[width=0.48\columnwidth]{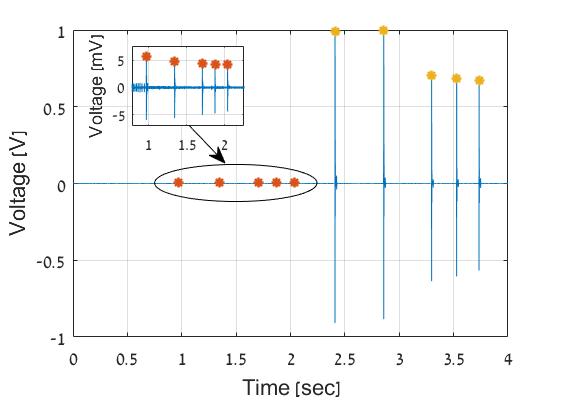}}
	\hspace{2mm}
 	\subfloat[\textbf{\textit{Cocktail-party}}: example of far-field annotation of three overlapping codas, each marked with a different color.
        \label{fig:Example2}]{\includegraphics[width=0.48\columnwidth]{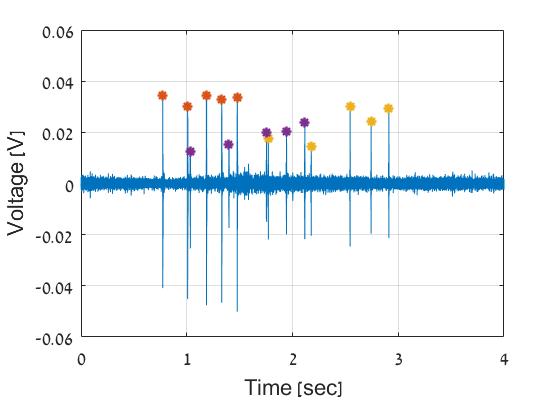}}
  	\hspace{2mm}
 	\subfloat[\textbf{\textit{High interference}}: example of far-field annotation of a coda signal in the presence of echolocation clicks of similar intensity. Red crosses mark the detected coda clicks.
        \label{fig:Example3}]{\includegraphics[width=0.48\columnwidth]{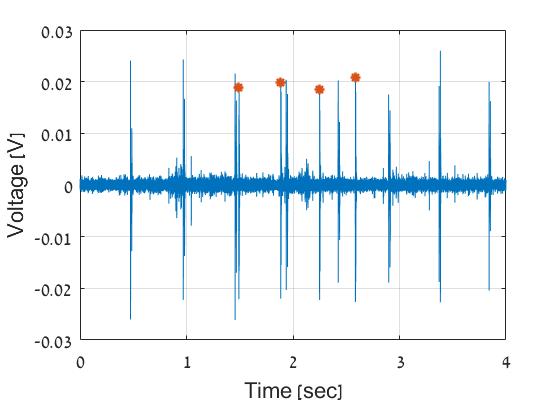}}
    \caption{Examples for the operation of the coda detector and annotator in three challenging scenarios.}
    \label{Toy examples}
\end{figure}

Examples of the operation of the detector and annotator are shown in Fig.~\ref{Toy examples}. Fig.~\ref{fig:Example1} shows the ability to detect both focal and non-focal whales in a wide dynamic range, where the intensity of the coda signal of the focal whale (red markers) is $200$ times stronger than that of the non-focal whale (yellow markers). An example of the annotator's ability to separate overlapping codas can be seen in Fig.~\ref{fig:Example2}, where three overlapping codas with similar SNR can be seen. Fig.~\ref{fig:Example1} shows an example of the separation of a coda signal from an echolocation signal with similar acoustic intensity.

%\bibliography{IEEEfull,BiomimickedBib}
\bibliography{IEEEfull,Codabibliography}

\begin{thebibliography}{10}
\urlstyle{rm}
\expandafter\ifx\csname url\endcsname\relax
  \def\url#1{\texttt{#1}}\fi
\expandafter\ifx\csname urlprefix\endcsname\relax\def\urlprefix{URL }\fi
\expandafter\ifx\csname doiprefix\endcsname\relax\def\doiprefix{DOI: }\fi
\providecommand{\bibinfo}[2]{#2}
\providecommand{\eprint}[2][]{\url{#2}}

\bibitem{andreas2021cetacean}
\bibinfo{author}{Andreas, J.} \emph{et~al.}
\newblock \bibinfo{journal}{\bibinfo{title}{Towards understanding the
  communication in sperm whales}}.
\newblock {\emph{\JournalTitle{i{S}cience}}} \textbf{\bibinfo{volume}{25}},
  \bibinfo{pages}{104393}, \doiprefix\url{10.1016/j.isci.2022.104393}
  (\bibinfo{year}{2022}).

\bibitem{goldwasser2023}
\bibinfo{author}{Goldwasser, S.}, \bibinfo{author}{Gruber, D.},
  \bibinfo{author}{Kalai, A.~T.} \& \bibinfo{author}{Paradise, O.}
\newblock \bibinfo{title}{A theory of unsupervised translation motivated by
  understanding animal communication}.
\newblock In \emph{\bibinfo{booktitle}{Thirty-seventh Conference on Neural
  Information Processing Systems}} (\bibinfo{year}{2023}).

\bibitem{muller2023soundscapes}
\bibinfo{author}{M{\"u}ller, J.} \emph{et~al.}
\newblock \bibinfo{journal}{\bibinfo{title}{Soundscapes and deep learning
  enable tracking biodiversity recovery in tropical forests}}.
\newblock {\emph{\JournalTitle{Nature communications}}}
  \textbf{\bibinfo{volume}{14}}, \bibinfo{pages}{6191} (\bibinfo{year}{2023}).

\bibitem{brakes2019animal}
\bibinfo{author}{Brakes, P.} \emph{et~al.}
\newblock \bibinfo{journal}{\bibinfo{title}{Animal cultures matter for
  conservation}}.
\newblock {\emph{\JournalTitle{Science}}} \textbf{\bibinfo{volume}{363}},
  \bibinfo{pages}{1032--1034} (\bibinfo{year}{2019}).

\bibitem{brakes2021deepening}
\bibinfo{author}{Brakes, P.} \emph{et~al.}
\newblock \bibinfo{journal}{\bibinfo{title}{A deepening understanding of animal
  culture suggests lessons for conservation}}.
\newblock {\emph{\JournalTitle{Proceedings of the Royal Society B}}}
  \textbf{\bibinfo{volume}{288}}, \bibinfo{pages}{20202718}
  (\bibinfo{year}{2021}).

\bibitem{rendell2003vocal}
\bibinfo{author}{Rendell, L.~E.} \& \bibinfo{author}{Whitehead, H.}
\newblock \bibinfo{journal}{\bibinfo{title}{Vocal clans in sperm whales
  ({{Physeter}} macrocephalus)}}.
\newblock {\emph{\JournalTitle{Proc. Royal Soc. B}}}
  \textbf{\bibinfo{volume}{270}}, \bibinfo{pages}{225--231},
  \doiprefix\url{10.1098/rspb.2002.2239} (\bibinfo{year}{2003}).

\bibitem{hersh2021method}
\bibinfo{author}{Hersh, T.~A.}, \bibinfo{author}{Gero, S.},
  \bibinfo{author}{Rendell, L.} \& \bibinfo{author}{Whitehead, H.}
\newblock \bibinfo{journal}{\bibinfo{title}{Using identity calls to detect
  structure in acoustic datasets}}.
\newblock {\emph{\JournalTitle{Methods in Ecology and Evolution}}}
  \textbf{\bibinfo{volume}{12}}, \bibinfo{pages}{1668--1678}
  (\bibinfo{year}{2021}).

\bibitem{sanchez2010efficient}
\bibinfo{author}{S{\'a}nchez-Garc{\'\i}a, A.}, \bibinfo{author}{Bueno-Crespo,
  A.} \& \bibinfo{author}{Sancho-G{\'o}mez, J.}
\newblock \bibinfo{journal}{\bibinfo{title}{An efficient statistics-based
  method for the automated detection of sperm whale clicks}}.
\newblock {\emph{\JournalTitle{Applied acoustics}}}
  \textbf{\bibinfo{volume}{71}}, \bibinfo{pages}{451--459}
  (\bibinfo{year}{2010}).

\bibitem{roch2008comparison}
\bibinfo{author}{Roch, M.~A.}, \bibinfo{author}{Soldevilla, M.~S.},
  \bibinfo{author}{Hoenigman, R.}, \bibinfo{author}{Wiggins, S.~M.} \&
  \bibinfo{author}{Hildebrand, J.~A.}
\newblock \bibinfo{journal}{\bibinfo{title}{Comparison of machine learning
  techniques for the classification of echolocation clicks from three species
  of odontocetes}}.
\newblock {\emph{\JournalTitle{Canadian Acoustics}}}
  \textbf{\bibinfo{volume}{36}}, \bibinfo{pages}{41--47}
  (\bibinfo{year}{2008}).

\bibitem{lohrasbipeydeh2014adaptive}
\bibinfo{author}{Lohrasbipeydeh, H.}, \bibinfo{author}{Dakin, D.~T.},
  \bibinfo{author}{Gulliver, T.~A.}, \bibinfo{author}{Amindavar, H.} \&
  \bibinfo{author}{Zielinski, A.}
\newblock \bibinfo{journal}{\bibinfo{title}{Adaptive energy-based acoustic
  sperm whale echolocation click detection}}.
\newblock {\emph{\JournalTitle{IEEE Journal of Oceanic Engineering}}}
  \textbf{\bibinfo{volume}{40}}, \bibinfo{pages}{957--968}
  (\bibinfo{year}{2014}).

\bibitem{kandia2006detection}
\bibinfo{author}{Kandia, V.} \& \bibinfo{author}{Stylianou, Y.}
\newblock \bibinfo{journal}{\bibinfo{title}{Detection of sperm whale clicks
  based on the teager--kaiser energy operator}}.
\newblock {\emph{\JournalTitle{Applied Acoustics}}}
  \textbf{\bibinfo{volume}{67}}, \bibinfo{pages}{1144--1163}
  (\bibinfo{year}{2006}).

\bibitem{kandia2008phase}
\bibinfo{author}{Kandia, V.} \& \bibinfo{author}{Stylianou, Y.}
\newblock \bibinfo{title}{A phase based detector of whale clicks}.
\newblock In \emph{\bibinfo{booktitle}{2008 New Trends for Environmental
  Monitoring Using Passive Systems}}, \bibinfo{pages}{1--6}
  (\bibinfo{organization}{IEEE}, \bibinfo{year}{2008}).

\bibitem{beslin2018automatic}
\bibinfo{author}{Beslin, W.~A.}, \bibinfo{author}{Whitehead, H.} \&
  \bibinfo{author}{Gero, S.}
\newblock \bibinfo{journal}{\bibinfo{title}{Automatic acoustic estimation of
  sperm whale size distributions achieved through machine recognition of
  on-axis clicks}}.
\newblock {\emph{\JournalTitle{The Journal of the Acoustical Society of
  America}}} \textbf{\bibinfo{volume}{144}}, \bibinfo{pages}{3485--3495}
  (\bibinfo{year}{2018}).

\bibitem{madsen2002sperm}
\bibinfo{author}{Madsen, P.} \emph{et~al.}
\newblock \bibinfo{journal}{\bibinfo{title}{Sperm whale sound production
  studied with ultrasound time/depth-recording tags}}.
\newblock {\emph{\JournalTitle{Journal of Experimental Biology}}}
  \textbf{\bibinfo{volume}{205}}, \bibinfo{pages}{1899--1906}
  (\bibinfo{year}{2002}).

\bibitem{Mohl2003}
\bibinfo{author}{M{\o}hl, B.}, \bibinfo{author}{Wahlberg, M.},
  \bibinfo{author}{Madsen, P.~T.}, \bibinfo{author}{Heerfordt, A.} \&
  \bibinfo{author}{Lund, A.}
\newblock \bibinfo{journal}{\bibinfo{title}{The monopulsed nature of sperm
  whale clicks}}.
\newblock {\emph{\JournalTitle{J. Acoust. Soc. Am.}}}
  \textbf{\bibinfo{volume}{114}}, \bibinfo{pages}{1143--1154}
  (\bibinfo{year}{2003}).

\bibitem{watkins1977sperm}
\bibinfo{author}{Watkins, W.~A.} \& \bibinfo{author}{Schevill, W.~E.}
\newblock \bibinfo{journal}{\bibinfo{title}{Sperm whale codas}}.
\newblock {\emph{\JournalTitle{The Journal of the Acoustical Society of
  America}}} \textbf{\bibinfo{volume}{62}}, \bibinfo{pages}{1485--1490}
  (\bibinfo{year}{1977}).

\bibitem{schulz2008overlapping}
\bibinfo{author}{Schulz, T.~M.}, \bibinfo{author}{Whitehead, H.},
  \bibinfo{author}{Gero, S.} \& \bibinfo{author}{Rendell, L.}
\newblock \bibinfo{journal}{\bibinfo{title}{Overlapping and matching of codas
  in vocal interactions between sperm whales: insights into communication
  function}}.
\newblock {\emph{\JournalTitle{Animal Behaviour}}}
  \textbf{\bibinfo{volume}{76}}, \bibinfo{pages}{1977--1988}
  (\bibinfo{year}{2008}).

\bibitem{weilgart1993coda}
\bibinfo{author}{Weilgart, L.} \& \bibinfo{author}{Whitehead, H.}
\newblock \bibinfo{journal}{\bibinfo{title}{Coda communication by sperm whales
  (physeter macrocephalus) off the galapagos islands}}.
\newblock {\emph{\JournalTitle{Canadian Journal of Zoology}}}
  \textbf{\bibinfo{volume}{71}}, \bibinfo{pages}{744--752}
  (\bibinfo{year}{1993}).

\bibitem{rendell2004shared}
\bibinfo{author}{Rendell, L.} \& \bibinfo{author}{Whitehead, H.}
\newblock \bibinfo{journal}{\bibinfo{title}{Do sperm whales share coda
  vocalizations? insights into coda usage from acoustic size measurement}}.
\newblock {\emph{\JournalTitle{Animal Behaviour}}}
  \textbf{\bibinfo{volume}{67}}, \bibinfo{pages}{865--874}
  (\bibinfo{year}{2004}).

\bibitem{rendell2003comparing}
\bibinfo{author}{Rendell, L.} \& \bibinfo{author}{Whitehead, H.}
\newblock \bibinfo{journal}{\bibinfo{title}{Comparing repertoires of sperm
  whale codas: a multiple methods approach}}.
\newblock {\emph{\JournalTitle{Bioacoustics}}} \textbf{\bibinfo{volume}{14}},
  \bibinfo{pages}{61--81} (\bibinfo{year}{2003}).

\bibitem{gero2016individual}
\bibinfo{author}{Gero, S.}, \bibinfo{author}{Whitehead, H.} \&
  \bibinfo{author}{Rendell, L.}
\newblock \bibinfo{journal}{\bibinfo{title}{Individual, unit and vocal clan
  level identity cues in sperm whale codas}}.
\newblock {\emph{\JournalTitle{Royal Society Open Science}}}
  \textbf{\bibinfo{volume}{3}}, \bibinfo{pages}{150372} (\bibinfo{year}{2016}).

\bibitem{Vachon2022oceanNomads}
\bibinfo{author}{Vachon, F.}, \bibinfo{author}{Hersh, T.~A.},
  \bibinfo{author}{Rendell, L.}, \bibinfo{author}{Gero, S.} \&
  \bibinfo{author}{Whitehead, H.}
\newblock \bibinfo{journal}{\bibinfo{title}{Ocean nomads or island specialists?
  culturally driven habitat partitioning contrasts in scale between
  geographically isolated sperm whale populations}}.
\newblock {\emph{\JournalTitle{Royal Society Open Science}}}
  \textbf{\bibinfo{volume}{9}}, \bibinfo{pages}{211737} (\bibinfo{year}{2022}).

\bibitem{sharma2023contextual}
\bibinfo{author}{Sharma, P.} \emph{et~al.}
\newblock \bibinfo{journal}{\bibinfo{title}{Contextual and combinatorial
  structure in sperm whale vocalisations}}.
\newblock {\emph{\JournalTitle{bioRxiv}}} \bibinfo{pages}{2023--12}
  (\bibinfo{year}{2023}).

\bibitem{huggenberger16}
\bibinfo{author}{Huggenberger, S.}, \bibinfo{author}{André, M.} \&
  \bibinfo{author}{Oelschläger, H. H.~A.}
\newblock \bibinfo{journal}{\bibinfo{title}{The nose of the sperm whale:
  overviews of functional design, structural homologies and evolution}}.
\newblock {\emph{\JournalTitle{Journal of the Marine Biological Association of
  the United Kingdom}}} \textbf{\bibinfo{volume}{96}},
  \bibinfo{pages}{783–806}, \doiprefix\url{10.1017/S0025315414001118}
  (\bibinfo{year}{2016}).

\bibitem{hersh2022evidence}
\bibinfo{author}{Hersh, T.~A.} \emph{et~al.}
\newblock \bibinfo{journal}{\bibinfo{title}{Evidence from sperm whale clans of
  symbolic marking in non-human cultures}}.
\newblock {\emph{\JournalTitle{Proc. Natl. Acad. Sci.}}}
  \textbf{\bibinfo{volume}{119}}, \bibinfo{pages}{e2201692119},
  \doiprefix\url{10.1073/pnas.2201692119} (\bibinfo{year}{2022}).

\bibitem{DtagPaper}
\bibinfo{author}{Johnson, M.~P.} \& \bibinfo{author}{Tyack, P.~L.}
\newblock \bibinfo{journal}{\bibinfo{title}{A digital acoustic recording tag
  for measuring the response of wild marine mammals to sound}}.
\newblock {\emph{\JournalTitle{IEEE Journal of Oceanic Engineering}}}
  \textbf{\bibinfo{volume}{28}}, \bibinfo{pages}{3--12} (\bibinfo{year}{2003}).

\bibitem{moca2021time}
\bibinfo{author}{Moca, V.~V.}, \bibinfo{author}{B{\^a}rzan, H.},
  \bibinfo{author}{Nagy-D{\u{a}}b{\^a}can, A.} \& \bibinfo{author}{Mureșan,
  R.~C.}
\newblock \bibinfo{journal}{\bibinfo{title}{Time-frequency super-resolution
  with superlets}}.
\newblock {\emph{\JournalTitle{Nature communications}}}
  \textbf{\bibinfo{volume}{12}}, \bibinfo{pages}{337} (\bibinfo{year}{2021}).

\bibitem{schulz2009off}
\bibinfo{author}{Schulz, T.~M.}, \bibinfo{author}{Whitehead, H.} \&
  \bibinfo{author}{Rendell, L.}
\newblock \bibinfo{journal}{\bibinfo{title}{Off-axis effects on the multi-pulse
  structure of sperm whale coda clicks}}.
\newblock {\emph{\JournalTitle{The Journal of the Acoustical Society of
  America}}} \textbf{\bibinfo{volume}{125}}, \bibinfo{pages}{1768--1773}
  (\bibinfo{year}{2009}).

\bibitem{jacobs2024active}
\bibinfo{author}{Jacobs, E.~R.} \emph{et~al.}
\newblock \bibinfo{journal}{\bibinfo{title}{The active space of sperm whale
  codas: inter-click information for intra-unit communication}}.
\newblock {\emph{\JournalTitle{Journal of Experimental Biology}}}
  \bibinfo{pages}{jeb--246442} (\bibinfo{year}{2024}).

\bibitem{mohl1981}
\bibinfo{author}{M{\o}hl, B.}, \bibinfo{author}{Larsen, E.} \&
  \bibinfo{author}{Amundin, M.}
\newblock \bibinfo{journal}{\bibinfo{title}{Sperm whale size determination:
  Outlines of an acoustic approach}}.
\newblock {\emph{\JournalTitle{Mammals in the Seas. Fisheries Series No. 5,
  Mammals of the Seas}}} \textbf{\bibinfo{volume}{3}},
  \bibinfo{pages}{327--331} (\bibinfo{year}{1981}).

\bibitem{gubnitsky2023detecting}
\bibinfo{author}{Gubnitsky, G.} \& \bibinfo{author}{Diamant, R.}
\newblock \bibinfo{journal}{\bibinfo{title}{Detecting the presence of sperm
  whales echolocation clicks in noisy environments}}.
\newblock {\emph{\JournalTitle{arXiv preprint arXiv:2401.00900}}}
  (\bibinfo{year}{2023}).

\bibitem{VachonAbundance}
\bibinfo{author}{Vachon, F.}, \bibinfo{author}{Rendell, L.},
  \bibinfo{author}{Gero, S.} \& \bibinfo{author}{Whitehead, H.}
\newblock \bibinfo{journal}{\bibinfo{title}{Abundance estimate of eastern
  caribbean sperm whales using large scale regional surveys}}.
\newblock {\emph{\JournalTitle{Marine Mammal Science}}}
  \doiprefix\url{https://doi.org/10.1111/mms.13116}.

\bibitem{Gero2007}
\bibinfo{author}{Gero, S.}, \bibinfo{author}{Gordon, J.},
  \bibinfo{author}{Carlson, C.}, \bibinfo{author}{Evans, P.} \&
  \bibinfo{author}{Whitehead, H.}
\newblock \bibinfo{journal}{\bibinfo{title}{Population estimate and
  inter-island movement of sperm whales, physeter macrocephalus, in the eastern
  caribbean sea}}.
\newblock {\emph{\JournalTitle{J. Cetacean Res. Manage.}}}
  \textbf{\bibinfo{volume}{9}}, \bibinfo{pages}{143--150}
  (\bibinfo{year}{2007}).

\bibitem{Gero2014}
\bibinfo{author}{Gero, S.} \emph{et~al.}
\newblock \bibinfo{journal}{\bibinfo{title}{Behavior and social structure of
  the sperm whales of dominica, west indies}}.
\newblock {\emph{\JournalTitle{Marine Mammal Science}}}
  \textbf{\bibinfo{volume}{30}}, \bibinfo{pages}{905--922}
  (\bibinfo{year}{2014}).

\bibitem{Gero2016-social}
\bibinfo{author}{Gero, S.}, \bibinfo{author}{B{\o}ttcher, A.},
  \bibinfo{author}{Whitehead, H.} \& \bibinfo{author}{Madsen, P.~T.}
\newblock \bibinfo{journal}{\bibinfo{title}{Socially segregated, sympatric
  sperm whale clans in the atlantic ocean}}.
\newblock {\emph{\JournalTitle{R Soc Open Sci}}} \textbf{\bibinfo{volume}{3}},
  \bibinfo{pages}{160061} (\bibinfo{year}{2016}).

\bibitem{schulz2011individual}
\bibinfo{author}{Schulz, T.~M.}, \bibinfo{author}{Whitehead, H.},
  \bibinfo{author}{Gero, S.} \& \bibinfo{author}{Rendell, L.}
\newblock \bibinfo{journal}{\bibinfo{title}{Individual vocal production in a
  sperm whale (physeter macrocephalus) social unit}}.
\newblock {\emph{\JournalTitle{Marine Mammal Science}}}
  \textbf{\bibinfo{volume}{27}}, \bibinfo{pages}{149--166}
  (\bibinfo{year}{2011}).

\bibitem{cheng2016listen}
\bibinfo{author}{Cheng, J.~T.}, \bibinfo{author}{Tracy, J.~L.},
  \bibinfo{author}{Ho, S.} \& \bibinfo{author}{Henrich, J.}
\newblock \bibinfo{journal}{\bibinfo{title}{Listen, follow me: Dynamic vocal
  signals of dominance predict emergent social rank in humans.}}
\newblock {\emph{\JournalTitle{Journal of Experimental Psychology: General}}}
  \textbf{\bibinfo{volume}{145}}, \bibinfo{pages}{536} (\bibinfo{year}{2016}).

\bibitem{kavanagh2021dominance}
\bibinfo{author}{Kavanagh, E.} \emph{et~al.}
\newblock \bibinfo{journal}{\bibinfo{title}{Dominance style is a key predictor
  of vocal use and evolution across nonhuman primates}}.
\newblock {\emph{\JournalTitle{Royal Society open science}}}
  \textbf{\bibinfo{volume}{8}}, \bibinfo{pages}{210873} (\bibinfo{year}{2021}).

\bibitem{arnbom1987}
\bibinfo{author}{Arnbom, T.}
\newblock \bibinfo{title}{Individual photographic identification : a key to the
  social organization of sperm whales} (\bibinfo{year}{1987}).

\bibitem{Webside_coda}
\bibinfo{author}{Gubnisky, G.}
\newblock \bibinfo{title}{Link to our matlab implementation code.}
  (\bibinfo{year}{2024}).
\newblock
  \bibinfo{note}{\url{https://drive.google.com/drive/folders/1RvymG3zkM7nFbWIEiSsBm5CGgpNy9xDw}.}

\bibitem{gubnitsky2023inter}
\bibinfo{author}{Gubnitsky, G.} \& \bibinfo{author}{Diamant, R.}
\newblock \bibinfo{title}{Inter-pulse estimation for sperm whale click
  detection}.
\newblock In \emph{\bibinfo{booktitle}{ICASSP 2023-2023 IEEE International
  Conference on Acoustics, Speech and Signal Processing (ICASSP)}},
  \bibinfo{pages}{1--5} (\bibinfo{organization}{IEEE}, \bibinfo{year}{2023}).

\bibitem{kaiser1990simple}
\bibinfo{author}{Kaiser, J.~F.}
\newblock \bibinfo{title}{On a simple algorithm to calculate the'energy'of a
  signal}.
\newblock In \emph{\bibinfo{booktitle}{International conference on acoustics,
  speech, and signal processing}}, \bibinfo{pages}{381--384}
  (\bibinfo{organization}{IEEE}, \bibinfo{year}{1990}).

\bibitem{pavan2000time}
\bibinfo{author}{Pavan, G.} \emph{et~al.}
\newblock \bibinfo{journal}{\bibinfo{title}{Time patterns of sperm whale codas
  recorded in the mediterranean sea 1985--1996}}.
\newblock {\emph{\JournalTitle{The Journal of the Acoustical Society of
  America}}} \textbf{\bibinfo{volume}{107}}, \bibinfo{pages}{3487--3495}
  (\bibinfo{year}{2000}).

\bibitem{amano2014differences}
\bibinfo{author}{Amano, M.}, \bibinfo{author}{Kourogi, A.},
  \bibinfo{author}{Aoki, K.}, \bibinfo{author}{Yoshioka, M.} \&
  \bibinfo{author}{Mori, K.}
\newblock \bibinfo{journal}{\bibinfo{title}{Differences in sperm whale codas
  between two waters off japan: possible geographic separation of vocal
  clans}}.
\newblock {\emph{\JournalTitle{Journal of Mammalogy}}}
  \textbf{\bibinfo{volume}{95}}, \bibinfo{pages}{169--175}
  (\bibinfo{year}{2014}).

\bibitem{huggenberger2014acoustic}
\bibinfo{author}{Huggenberger, S.}, \bibinfo{author}{Andr{\'e}, M.} \&
  \bibinfo{author}{Oelschl{\"a}ger, H.~H.}
\newblock \bibinfo{journal}{\bibinfo{title}{An acoustic valve within the nose
  of sperm whales p hyseter macrocephalus}}.
\newblock {\emph{\JournalTitle{Mammal review}}} \textbf{\bibinfo{volume}{44}},
  \bibinfo{pages}{81--87} (\bibinfo{year}{2014}).

\bibitem{garey1997computers}
\bibinfo{author}{Garey, M.~R.}
\newblock \bibinfo{journal}{\bibinfo{title}{Computers and intractability: A
  guide to the theory of np-completeness, freeman}}.
\newblock {\emph{\JournalTitle{Fundamental}}}  (\bibinfo{year}{1997}).

\bibitem{gordon1991evaluation}
\bibinfo{author}{Gordon, J.~C.}
\newblock \bibinfo{journal}{\bibinfo{title}{Evaluation of a method for
  determining the length of sperm whales (physeter catodon) from their
  vocalizations}}.
\newblock {\emph{\JournalTitle{Journal of Zoology}}}
  \textbf{\bibinfo{volume}{224}}, \bibinfo{pages}{301--314}
  (\bibinfo{year}{1991}).

\bibitem{diamant2012and}
\bibinfo{author}{Diamant, R.}, \bibinfo{author}{Tan, H.-P.} \&
  \bibinfo{author}{Lampe, L.}
\newblock \bibinfo{journal}{\bibinfo{title}{Los and nlos classification for
  underwater acoustic localization}}.
\newblock {\emph{\JournalTitle{IEEE Transactions on mobile Computing}}}
  \textbf{\bibinfo{volume}{13}}, \bibinfo{pages}{311--323}
  (\bibinfo{year}{2012}).

\bibitem{kotz1975multivariate}
\bibinfo{author}{Kotz, S.}
\newblock \bibinfo{title}{Multivariate distributions at a cross road}.
\newblock In \emph{\bibinfo{booktitle}{A Modern Course on Statistical
  Distributions in Scientific Work: Volume 1—Models and Structures
  Proceedings of the NATO Advanced Study Institute held at the University of
  Calgagry, Calgary, Alberta, Canada July 29--August 10, 1974}},
  \bibinfo{pages}{247--270} (\bibinfo{publisher}{Springer},
  \bibinfo{year}{1975}).

\bibitem{boukouvalas2015new}
\bibinfo{author}{Boukouvalas, Z.}, \bibinfo{author}{Said, S.},
  \bibinfo{author}{Bombrun, L.}, \bibinfo{author}{Berthoumieu, Y.} \&
  \bibinfo{author}{Adal{\i}, T.}
\newblock \bibinfo{journal}{\bibinfo{title}{A new riemannian averaged
  fixed-point algorithm for mggd parameter estimation}}.
\newblock {\emph{\JournalTitle{IEEE Signal Processing Letters}}}
  \textbf{\bibinfo{volume}{22}}, \bibinfo{pages}{2314--2318}
  (\bibinfo{year}{2015}).

\bibitem{najar2018unsupervised}
\bibinfo{author}{Najar, F.}, \bibinfo{author}{Bourouis, S.},
  \bibinfo{author}{Zaguia, A.}, \bibinfo{author}{Bouguila, N.} \&
  \bibinfo{author}{Belghith, S.}
\newblock \bibinfo{title}{Unsupervised human action categorization using a
  riemannian averaged fixed-point learning of multivariate ggmm}.
\newblock In \emph{\bibinfo{booktitle}{Image Analysis and Recognition: 15th
  International Conference, ICIAR 2018, P{\'o}voa de Varzim, Portugal, June
  27--29, 2018, Proceedings 15}}, \bibinfo{pages}{408--415}
  (\bibinfo{organization}{Springer}, \bibinfo{year}{2018}).

\bibitem{schwarz1978estimating}
\bibinfo{author}{Schwarz, G.}
\newblock \bibinfo{journal}{\bibinfo{title}{Estimating the dimension of a
  model}}.
\newblock {\emph{\JournalTitle{The annals of statistics}}}
  \bibinfo{pages}{461--464} (\bibinfo{year}{1978}).

\end{thebibliography}

\vspace{0.25cm}

\section*{Acknowledgements}

This study was funded by Project CETI via grants from Dalio Philanthropies and Ocean X; Sea Grape Foundation; Virgin Unite, Rosamund Zander/Hansjorg Wyss, Chris Anderson/Jacqueline Novogratz through The Audacious Project: a collaborative funding initiative housed at TED. We thank the Chief Fisheries Officers and the Dominica Fisheries Division officers for research permits and their collaboration in data collection; all the crews of R/V Balaena and The DSWP team for data collection, curation, and annotation of audio recordings across 15 years; as well as Dive Dominica, Al Dive, and W.E.T. Dominica for logistical support while in Dominica. We are grateful to Kristian Beedholm for CodaSorter, as well as Mark Johnson and Peter L. Tyack for their in-kind contribution of Dtags and associated code during DSWP field research in 2014-2018.

\section*{Availability of data and materials}

 Code availability of the implementation is available in the link\\
\url{https://drive.google.com/drive/folders/1RvymG3zkM7nFbWIEiSsBm5CGgpNy9xDw}~\cite{Webside_coda}.

\section*{Author contributions statement}

G.G conceived the detector and annotator design and processed the data. Y.M. assisted in CETI data collection and manual annotation.
S.G conducted and funded DSWP fieldwork, collected, annotated, and curated DSWP dataset, assisted in CETI data collection, edited the manuscript, and assisted in deriving conclusions.
D.G provided funding and management and assisted in deriving conclusions. R.D conceived the detector and annotator design, assisted in CETI data collection and was in charge of manuscript write up.

\section*{Additional information}

The authors declare no conflicts of interest.

% The corresponding author is responsible for submitting a \href{http://www.nature.com/srep/policies/index.html#competing}{competing interests statement} on behalf of all authors of the paper. This statement must be included in the submitted article file.

%\clearpage

%\input{rebuttal_letter_01}

\end{document}